\documentclass[11pt]{article}
\usepackage{amsmath,amssymb,graphicx}

\newcommand{\ba}{\begin{alignat}{3}}
\newcommand{\e}{\epsilon}
\newcommand{\p}{\partial}
\newcommand{\dl}{\delta}

\newcommand{\g}{\gamma}

\newcommand{\pa}{\partial}

\newcommand{\om}{\omega}

\begin{document}

\begin{titlepage}
\begin{flushright}
\end{flushright}
\begin{center}
  \vspace{4cm}
  {\bf \Large Quantum Near Horizon Geometry of Black 0-Brane}
  \\  \vspace{2cm}
  Yoshifumi Hyakutake
   \\ \vspace{1cm}
   {\it College of Science, Ibaraki University \\
   Bunkyo 1-1, Mito, Ibaraki 310-0062, Japan}
\end{center}

\vspace{2cm}
\begin{abstract}
We investigate a bunch of D0-branes to reveal its quantum nature from the gravity side.
In the classical limit, it is well described by a non-extremal black 0-brane in type IIA supergravity.
The solution is uplifted to the eleven dimensions and expressed by a non-extremal M-wave solution.
After reviewing the effective action for the M-theory, we explicitly solve the equations of motion for 
the near horizon geometry of the M-wave.
As a result we derive an unique solution which includes the effect of the quantum gravity.
Thermodynamic property of the quantum near horizon geometry of the black 0-brane is also
studied by using Wald's formula. Combining our result with that of the Monte Carlo simulation
of the dual thermal gauge theory, we find strong evidence for the gauge/gravity duality in the 
D0-branes system at the level of quantum gravity.
\end{abstract}
\end{titlepage}

\setlength{\baselineskip}{0.65cm}

\section{Introduction}

Superstring theory is a promising candidate for the theory of quantum gravity,
and it plays important roles to reveal quantum nature of black holes.
Fundamental objects in the superstring theory are D-branes as well as strings~\cite{Pol},
and in the low energy limit their dynamics are governed by supergravity.
The D-branes are described by classical solutions in the supergravity, 
which are called black branes~\cite{GM,HS}.
A special class of them has event horizon like the black holes and its entropy can be evaluated by the area law.
Interestingly the entropy can be statistically explained by counting number of microstates 
in the gauge theory on the D-branes~\cite{SV}.
This motivates us to study the black hole thermodynamics from the gauge theory.
Furthermore it is conjectured that the near horizon geometry of the black brane corresponds
to the gauge theory on the D-branes~\cite{M}. 
If this gauge/gravity duality is correct, 
the strong coupling limit of the gauge theory can be analyzed by the supergravity~\cite{GKP,W2}.

In this paper we consider a bunch of D0-branes in type IIA superstring theory.
In the low energy limit, a bunch of D0-branes with additional internal energy are well described 
by non-extremal black 0-brane solution in type IIA supergravity~\cite{GM,HS}.
After taking near horizon limit, the metric becomes AdS black hole like geometry 
in ten dimensional space-time~\cite{IMSY}.
From the gauge/gravity duality, this geometry corresponds to the strong coupling limit of
the gauge theory on the D0-branes~\cite{IMSY}, which is described by (1+0)-dimensional 
$U(N)$ super Yang-Mills theory~\cite{W}.
This gauge theory is paid much attention as nonperturbative definition of M-theory~\cite{dWHN,BFSS}, 
which is the strong coupling description of the type IIA superstring theory~\cite{T,W3}.
Recently nonperturbative aspects of the gauge theory are studied by the computer simulation~\cite{KLL}-\cite{HHIN}.
(See refs. \cite{CKU}, \cite{N} for reviews including other topics.)
Especially in ref.~\cite{HHNT}, physical quantities of the thermal gauge theory, such as the internal energy, 
are evaluated numerically, and a direct test of the gauge/gravity duality is performed including
$\alpha'$ correction to the type IIA supergravity.
Furthermore, if the internal energy of the black 0-brane can be evaluated precisely from the gravity side
including $g_s$ correction,
it is possible to give a direct test for the gauge/gravity duality
at the level of quantum gravity~\cite{HHIN}.
($\alpha'=\ell_s^2$ is the string length squared and $g_s$ is the string coupling constant.)

The purpose of this paper is to derive quantum correction to the near horizon geometry of 
the non-extremal black 0-brane directly from the gravity side. 
In order to do this, we need to know an effective action which include
quantum correction to the type IIA supergravity. In principle the effective action can be constructed
so as to be consistent with the scattering amplitudes in the type IIA superstring theory~\cite{GW},
and it is expressed by double expansion of $\alpha'$ and $g_s$.
For example, since four point amplitudes of gravitons at tree and one loop level are nontrivial,
there should exist terms like $\alpha'^3 e^{-2\phi} t_8 t_8 R^4$ and $\alpha'^3 g_s^2 t_8 t_8 R^4$ 
in the effective action, respectively~\cite{GW}--\cite{H}. 
These are called higher derivative terms and 
$t_8$ represents products of four Kronecker's deltas with eight indices.
Especially we are interested in the latter terms, which give nontrivial $g_s$ corrections to the geometry.
These higher derivative terms often play important roles to count the entropy of 
extremal black holes~\cite{D,OSV}.

It is necessary that the effective action of the type IIA superstring should possess local supersymmetry
in ten dimensions.
So the supersymmetrization of $\alpha'^3 g_s^2 t_8 t_8 R^4$ is very important~\cite{RSW1,RSW2,Su,PVW,HO,H}
to understand the structure of effective action.
Although the task is not completed yet, since our interest is on the geometry of the black 0-brane, 
it is enough to know terms which contain the metric, dilaton field and R-R 1-form field only.
Notice that these fields are collected into the metric in eleven dimensional supergravity~\cite{CJS},
and the black 0-brane is expressed by M-wave solution.
Then $\alpha'^3 g_s^2 t_8 t_8 R^4$ and other terms which include the dilaton and R-R 1-form field
are simply collected into $\ell_p^6 t_8t_8 R^4$ terms in eleven dimensions.
Here $\ell_p = \ell_s g_s^{1/3}$ is the Planck length in eleven dimensions.
Thus we consider the effective action for the M-theory and investigate quantum corrections
to the near horizon geometry of the non-extremal M-wave.
We show equations of motion for the effective action and explicitly solve them up to the order of $g_s^2$.
The M-wave geometry receives the quantum corrections and thermodynamic quantities for the M-wave are modified.
Especially the internal energy of the M-wave is obtained quantitatively including quantum effect
of the gravity. 

Organization of the paper is as follows.
In section 2, we review the classical near horizon geometry of the black 0-brane in ten dimensions, 
and uplift it to that of the M-wave in eleven dimensions.
In section 3, we discuss the higher derivative corrections in the type IIA superstring theory and the M-theory,
and solve the equations of motion for the near horizon geometry of the non-extremal M-wave in section 4.
In section 5, we evaluate the entropy and the energy of the M-wave up to $1/N^2$.
We probe the quantum near horizon geometry by D0-brane in section 6 and clarify the validity
of our analyses in section 7.
Section 8 is devoted to conclusion and discussion.
Detailed calculations and discussions on the ambiguities of the higher derivative corrections
are collected in the appendices.

\section{Classical Near Horizon Geometry of Black 0-Brane}

In this section, we briefly review the non-extremal solution of the black 0-brane
which carries mass and R-R charge.
Especially we uplift the solution to eleven dimensions and show that the black 0-brane
is described by the M-wave solution.

In the low energy limit, the dynamics of massless modes in type IIA superstring theory 
are governed by type IIA supergravity.
Since we are interested in the black 0-brane which couples to the graviton $g_{\mu\nu}$, the dilaton $\phi$
and R-R 1-form field $C_\mu$, the relevant part of the type IIA supergravity action is given by
\ba
  S_{10}^{(0)} &= \frac{1}{2\kappa_{10}^2} \int d^{10} x \sqrt{-g} 
  \Big\{ e^{-2\phi} \big( R + 4 \p_\mu \phi \p^\mu \phi \big) - \frac{1}{4} G_{\mu\nu} G^{\mu\nu} \Big\},
  \label{eq:10dsugra}
\end{alignat}
where $2\kappa_{10}^2 = (2\pi)^7 \ell_s^8 g_s^2$ and $G_{\mu\nu}$ is the field strength of $C_\mu$.
$g_s$ and $\ell_s$ are the string coupling constant and the string length, respectively.
It is possible to solve the equations of motion by making the ansatz 
that the metric is static and has SO(9) rotation symmetry.
Then we obtain non-extremal solution of the black 0-brane. (See ref.~\cite{AGMOO} for example.)
\begin{alignat}{3}
  &ds_{10}^2 = - \tilde{H}^{-\frac{1}{2}} \tilde{F} dt^2 
  + \tilde{H}^{\frac{1}{2}} \tilde{F}^{-1} dr^2 
  + \tilde{H}^\frac{1}{2} r^2 d\Omega_8^2 , \label{eq:sugrasol}
  \\
  &e^\phi = \tilde{H}^\frac{3}{4}, 
  \quad C = \Big(\frac{r_+}{r_-}\Big)^\frac{7}{2} \tilde{H}^{-1} dt, \notag
  \\[-0.1cm]
  &\tilde{H} = 1 + \frac{r_-^7}{r^7}, \quad \tilde{F} = 1 - \frac{r_+^7 - r_-^7}{r^7}. \notag
\end{alignat}
The horizon is located at $r_H = (r_+^7 - r_-^7)^{\frac{1}{7}}$.
Parameters $r_\pm$ are related to the mass $M_0$ and the R-R charge $Q_0$ of the black 0-brane by
\begin{alignat}{3}
  M_0 = \frac{V_{S^8}}{2\kappa_{10}^2} \big( 8 r_+^7 - r_-^7 \big), \qquad
  Q_0 = \frac{N}{\ell_s g_s} = \frac{7 V_{S^8}}{2\kappa_{10}^2} \big( r_+ r_- \big)^{\frac{7}{2}}, 
\end{alignat}
where $N$ is a number of D0-branes and 
$V_{S^8} = \frac{2 \pi^{9/2}}{\Gamma(9/2)} = \frac{2(2\pi)^4}{7\cdot 15}$ is the volume of $S^8$.
Now the parameters $r_\pm$ are expressed as
\begin{alignat}{3}
  r_\pm^7 &= (1+\delta)^{\pm 1}(2\pi)^2 15\pi g_s N \ell_s^7,
\end{alignat}
where $\delta$ is a non-negative parameter. The extremal limit $r_+=r_-$ is saturated when $\dl=0$.

Let us rewrite the solution (\ref{eq:sugrasol}) in terms of $U=r/\ell_s^2$ 
and $\lambda = g_s N / (2\pi)^2 \ell_s^3$, which correspond to typical energy scale and
't Hooft coupling in the dual gauge theory, respectively.
The near horizon limit of the non-extremal black 0-brane is taken by $\ell_s \to 0$ 
while $U$, $\lambda$ and $\delta /\ell_s^4$ are fixed.
Then the near horizon limit of the solution (\ref{eq:sugrasol}) becomes~\cite{IMSY}
\begin{alignat}{3}
  &ds_{10}^2 = \ell_s^2 \big( - H^{-\frac{1}{2}} F dt^2 + H^{\frac{1}{2}} F^{-1} dU^2
  + H^\frac{1}{2} U^2 d\Omega_8^2 \big) , \label{eq:nearH}
  \\
  &e^{\phi} = \ell_s^{-3} H^\frac{3}{4}, 
  \quad C = \ell_s^4 H^{-1} dt, \notag 
  \\
  &H = \frac{(2\pi)^4 15\pi \lambda}{U^7}, \quad 
  F = 1 - \frac{U_0^7}{U^7}, \notag
\end{alignat}
where $U_0^7 = \frac{2\delta}{\ell_s^4}(2\pi)^4 15\pi \lambda$.

The type IIA supergravity is related to the eleven dimensional supergravity via circle compactification.
In fact, the eleven dimensional metric is related to the ten dimensional one like
$ds_{11}^2 = e^{-2\phi/3} ds_{10}^2 + e^{4\phi/3} (dz - C_\mu dx^\mu)^2$.
The near horizon limit of the non-extremal solution of the black 0-brane (\ref{eq:nearH}) can 
be uplifted to eleven dimensions as
\begin{alignat}{3}
  ds_{11}^2 &= \ell_s^4 \big( - H^{-1} F dt^2 + F^{-1} dU^2 
  + U^2 d\Omega_8^2 + (\ell_s^{-4} H^{\frac{1}{2}} dz - H^{-\frac{1}{2}} dt)^2 \big). \label{eq:11nearH}
\end{alignat}
This represents the near horizon limit of the non-extremal M-wave solution in eleven dimensions.
The solution is purely geometrical and the expressions become simple.
Furthermore, on the geometrical part, quantum corrections to the eleven dimensional supergravity are under control.
This is the reason why we execute analyses of the solution in eleven dimensions.

\section{Quantum Correction to Eleven Dimensional Supergravity}

The eleven dimensional supergravity is realized as the low energy limit of the M-theory.
A fundamental object in the M-theory is a membrane and if we could take account of interaction of membranes, 
the effective action of the M-theory would become the eleven dimensional supergravity 
with some higher derivative terms.
Unfortunately quantization of the membrane has not been completed so far.
It is, however, possible to derive the relevant part of the quantum corrections in the M-theory 
by requiring local supersymmetry.
In this section we review the quantum corrections to the eleven dimensional supergravity.

Massless fields of the eleven dimensional supergravity consists of
a vielbein $e^a{}_\mu$, a Majorana gravitino $\psi_\mu$ and a 3-form field $A_{\mu\nu\rho}$.
Since we are only interested in the M-wave solution, we only need to take account of
the action which only depends on the graviton.
\ba
  2 \kappa_{11}^2 S^{(0)}_{11} &= \int d^{11}x \, e R, \label{eq:11dsugra}
\end{alignat}
where $2\kappa_{11}^2 = (2\pi)^8 \ell_p^9 = (2\pi)^8 \ell_s^9 g_s^3$.
Notice that after the dimensional reduction this becomes the action (\ref{eq:10dsugra}),
which contains the dilation and the R-R 1-form field as well as the graviton in ten dimensions~\cite{CJS}.

Of course there are other terms which depend on $\psi_\mu$ and $A_{\mu\nu\rho}$,
which are completely determined by the local supersymmetry.
For example, a variation of the vielbein under the local supersymmetry is 
given by $\dl [e] = [\overline{\e} \psi]$. 
Here we use a symbol $[X]$ to abbreviate indices and gamma matrices in $X$,
and $\epsilon$ represents a parameter of the local supersymmetry.
Then the variation of the scalar curvature is written by $\dl [e R] = [eR\overline{\e}\psi]$.
In order to cancel this, we see that a variation of the Majorana gravitino should 
include $\dl[\psi] = [D\e] + \cdots$ and simultaneously
there should exist a term like $[e\overline{\psi}\psi_2]$ in the action.
Here $\psi_2$ represents the field strength of the Majorana gravitino.
By continuing this process, it is possible to determine the structure of the 11 dimensional supergravity
completely~\cite{CJS}.

Now let us discuss quantum corrections to the eleven dimensional supergravity.
Since the M-theory is related to the type IIA superstring theory by the dimensional reduction,
the effective action of the M-theory should contain that of the type IIA superstring theory.
The latter can be obtained so as to be consistent with scattering amplitudes of strings,
and it is well-known that leading corrections to the type IIA supergravity include terms like $[eR^4]$.
This is directly uplifted to the eleven dimensions and we see that the effective action of the M-theory
should include terms like $B_1 = [e R^4]_7$. 
The subscript 7 indicates that there are potentially 7 independent terms
if we consider possible contractions of 16 indices out of 4 Riemann tensors.
(To be more precise, we excluded terms which contain Ricci tensor or scalar curvature,
since these can be eliminated by redefinition of the graviton. 
Discussions on these terms will be found in the appendix \ref{sec:R4}.)
As in the case of the eleven dimensional supergravity, it is possible to determine
other corrections by requiring the local supersymmetry.
For example, variations of $B_1$ under the local supersymmetry contain terms like $V_1 = [eR^4\bar{\epsilon}\psi]$.
In order to cancel these terms, $B_{11} = [e\epsilon_{11}AR^4]_{2}$ and 
$F_1 = [eR^3\bar{\psi}\psi_2]_{92}$ should exist in the action. 
The structures of $B_1$, $B_{11}$ and $F_1$ are severely restricted by the local supersymmetry.
By continuing this process, it is possible to show that a combination of terms in $B_1$ 
are completely determined up to over all factor~\cite{HO,H}. The result become as follows.
\begin{alignat}{3}
  2 \kappa_{11}^2 S_{11}^{(1)} &= 
  \frac{\pi^2\ell_p^6}{3\cdot 2^8 4!} \int d^{11}x \; e \Big(t_8 t_8 R^4 - \frac{1}{4!} \e_{11} \e_{11} R^4 \Big)
  \notag
  \\
  &= \frac{\pi^2\ell_p^6}{3\cdot 2^8 4!} \int d^{11}x \; e \Big\{
  24 \big( R_{abcd} R_{abcd} R_{efgh} R_{efgh}
  - 64 R_{abcd} R_{aefg} R_{bcdh} R_{efgh} \notag
  \\
  &\qquad\qquad\qquad\qquad\;\;
  + 2 R_{abcd} R_{abef} R_{cdgh} R_{efgh}
  + 16 R_{acbd} R_{aebf} R_{cgdh} R_{egfh} \notag
  \\
  &\qquad\qquad\qquad\qquad\;\;
  - 16 R_{abcd} R_{aefg} R_{befh} R_{cdgh}
  - 16 R_{abcd} R_{aefg} R_{bfeh} R_{cdgh} \big)
  \Big\}. \label{eq:R4}
\end{alignat}
Here $t_8$ is products of four Kronecker's deltas with eight indices
and $\e_{11}$ is an antisymmetric tensor with eleven indices.
Local Lorentz indices are labelled by $a,b, \cdots = 0,1,\cdots,10$.
Although all indices are lowered, it is understood those are contracted by the flat metric $\eta_{ab}$.
The Riemann tensor with local Lorentz indices is defined by
$R_{abcd} = e^\mu{}_c e^\nu{}_d (\pa_\mu \om_{\nu ab} - \pa_\nu \om_{\mu ab} 
+ \om_{\mu a}{}^e \om_{\nu eb} - \om_{\nu a}{}^e \om_{\mu eb})$, where
$\om_{\mu ab}$ is a spin connection and $\mu,\nu$ are space-time indices.
The over all factor in eq.~(\ref{eq:R4}) is determined by employing the result of 1-loop four graviton
amplitude in the type IIA superstring theory.

Since the near horizon limit of the M-wave solution (\ref{eq:11nearH}) is purely geometrical,
it is possible to examine the leading quantum corrections to it from the action (\ref{eq:R4}). 
Other terms which depend on the 3-form field are irrelevant to the analyses for the M-wave.
In summary the effective action of the M-theory is described by
\begin{alignat}{3}
  S_{11} = S_{11}^{(0)} + S_{11}^{(1)} =
  \frac{1}{2\kappa_{11}^2} \int d^{11}x \; e \Big\{ R
  + \gamma\ell_s^{12} \Big(t_8 t_8 R^4 - \frac{1}{4!} \e_{11} \e_{11} R^4 \Big) \Big\}, \label{eq:Maction}
\end{alignat}
where $\gamma = \frac{\pi^2}{3\cdot 2^8 4!} \frac{g_s^2}{\ell_s^6}
= \frac{\pi^6}{2^7 3^2} \frac{\lambda^2}{N^2}$. 
Notice that the parameter $\g$ remains finite after the decoupling limit is taken.
After the dimensional reduction, the action (\ref{eq:Maction}) becomes the effective action of 
the type IIA superstring theory, which includes the 1-loop effect of the gravity.

Now we derive equations of motion for the action (\ref{eq:Maction}).
Although the derivation is straightforward, we need to labor at many calculations 
because of the higher derivative terms in the action.
Therefore in practice we use the Mathematica code for the calculations.
Below we show the points of the calculations to build the code.

First of all we list variations of the fields with respect to the vielbein.
\begin{alignat}{3}
  &\delta e = - e e^i{}_\mu \delta e^\mu{}_i = - e \eta_{ij} \delta e^{ij}, \notag
  \\
  &\delta \omega_{cab} = e^\rho{}_c \dl \om_{\rho ab} = (\delta^k_{[a}\eta_{b]i}\eta_{cj} 
  + \delta^k_{[a} \eta_{b]j} \eta_{ci} + \delta^k_c \eta_{i[a} \eta_{b]j}) 
  D_k \delta e^{ij}, \notag
  \\
  &\delta R_{abcd} = \delta e^\mu{}_c R_{ab\mu d} + \delta e^\mu{}_d R_{abc\mu}
  + e^\mu{}_c e^\nu{}_d \delta R_{ab\mu\nu}
  = -2 \delta e^{ij} R_{abi[c} \eta_{d]j} + 2 D_{[c} \delta \omega_{d]ab}, \label{eq:fieldvar}
  \\
  &\delta R_{ab} 
  = - \delta e^{ij} R_{ajib} + \delta e^{ij} R_{ai} \eta_{bj} 
  + D_b \delta \omega^c{}_{ac} - D_c \delta \omega_{ba}{}^c, \notag
\end{alignat}
where $\dl e^{ij} \equiv e^i{}_\mu \dl e^{\mu j}$.
Then variations of the higher derivative terms are evaluated as
\ba
  &e \,\delta \Big(t_8 t_8 R^4 - \frac{1}{4!} \e_{11} \e_{11} R^4 \Big) \notag
  \\
  &= 24 e \big\{ 4 (\dl R_{abcd}) R_{abcd} R_{efgh} R_{efgh}
  - 64 (\dl R_{abcd}) R_{abce} R_{dfgh} R_{efgh} \notag
  \\
  &\qquad\quad
  + 8 (\dl R_{abcd}) R_{abef} R_{cdgh} R_{efgh}
  + 64 (\dl R_{abcd}) R_{aecg} R_{bfdh} R_{efgh} \notag
  \\
  &\qquad\quad
  - 64 (\dl R_{abcd}) R_{abeg} R_{cfeh} R_{dfgh} - 64 (\dl R_{abcd}) R_{efag} R_{efch} R_{gbhd} \notag
  \\
  &\qquad\quad
  + 32 (\dl R_{abcd}) R_{abef} R_{cegh} R_{dfgh} \big\} \notag
  \\
  &= e (\dl R_{abcd}) X_{abcd} \notag
  \\
  &= 2 e \delta e^{ij} R_{abci} X^{abc}{}_j 
  - 2 e X^{abcd} D_{d} \delta \omega_{cab} \notag
  \\
  &\cong 2 e \delta e^{ij} R_{abci} X^{abc}{}_j 
  - 2 e (\delta^k_{a}\eta_{bi}\eta_{cj} 
  + \delta^k_{a} \eta_{bj} \eta_{ci} + \delta^k_c \eta_{ia} \eta_{bj}) 
  D_k D_d X^{abcd} \delta e^{ij} \notag
  \\
  &= 2 e R_{abci} X^{abc}{}_j \delta e^{ij} 
  - 2 e D^c D^d (X_{cijd} + X_{cjid} + X_{ijcd}) \delta e^{ij} \notag
  \\
  &= e (3R_{abci} X^{abc}{}_j - R_{abcj}X^{abc}{}_i) \delta e^{ij}
  - 2 e D^c D^d (X_{cijd} + X_{cjid}) \delta e^{ij} \notag
  \\
  &= e \big(3R_{abci} X^{abc}{}_j - R_{abcj}X^{abc}{}_i
  - 4 D_{(a} D_{b)} X^a{}_{ij}{}^b \big) \delta e^{ij}, \label{eq:R4var}
\end{alignat}
where we defined
\ba
  X_{abcd} &= \frac{1}{2} \big( X'_{[ab][cd]} + X'_{[cd][ab]} \big), \label{eq:X}
  \\
  X'_{abcd} &= 96 \big(
  R_{abcd} R_{efgh} R_{efgh} - 16 R_{abce} R_{dfgh} R_{efgh} + 2 R_{abef} R_{cdgh} R_{efgh} \notag
  \\
  &\qquad\,
  + 16 R_{aecg} R_{bfdh} R_{efgh} - 16 R_{abeg} R_{cfeh} R_{dfgh} - 16 R_{efag} R_{efch} R_{gbhd} \notag
  \\
  &\qquad\,
  + 8 R_{abef} R_{cegh} R_{dfgh} \big) \notag.
\end{alignat}
Finally we obtain the equations of motion for the effective action (\ref{eq:Maction}).
\begin{alignat}{3}
  E_{ij} &\equiv R_{ij} - \frac{1}{2} \eta_{ij} R + \gamma\ell_s^{12} \Big\{
  - \frac{1}{2} \eta_{ij} \Big(t_8 t_8 R^4 - \frac{1}{4!} \e_{11} \e_{11} R^4 \Big) \notag
  \\
  &\quad
  + \frac{3}{2} R_{abci} X^{abc}{}_j - \frac{1}{2} R_{abcj}X^{abc}{}_i - 2 D_{(a} D_{b)} X^a{}_{ij}{}^b 
  \Big\} = 0. \label{eq:MEOM}
\end{alignat}
As mentioned before the action (\ref{eq:Maction}) is not unique
due to the ambiguity of field redefinitions, such as 
$g_{\mu\nu} \to g'_{\mu\nu} = g_{\mu\nu} + \g \ell_s^{12} R^2 R_{\mu\nu}$.
Therefore the equations of motion are not unique as well.
We will discuss, however, that physical quantities of the M-wave do not depend on
these ambiguities. (See appendix \ref{sec:app3}.)

\section{Quantum Near Horizon Geometry of Black 0-Brane}\label{sec:sol}

In the previous section, we have explained the effective action of the M-theory (\ref{eq:Maction}),
and derived the equations of motion (\ref{eq:MEOM}).
In this section we solve them up to the linear order of $\gamma$ and obtain the non-extremal
solution of the M-wave with quantum gravity correction.

In order to obtain the solution of (\ref{eq:MEOM}), 
we relax the ansatz for the M-wave as
\begin{alignat}{3}
  ds_{11}^2 &= \ell_s^4 \Big( - H_1^{-1} F_1 dt^2 + F_1^{-1} U_0^2 dx^2
  + U_0^2 x^2 d\Omega_8^2 + \big( \ell_s^{-4} H_2^{\frac{1}{2}} dz 
  - H_3^{-\frac{1}{2}} dt \big)^2 \Big), \label{eq:Mansatz}
  \\
  H_i &= \frac{(2\pi)^4 15\pi \lambda}{U_0^7} \Big(\frac{1}{x^7} + \frac{\gamma}{U_0^6} h_i\Big), \qquad
  F_1 = 1 - \frac{1}{x^7} + \frac{\gamma}{U_0^6} f_1, \notag
\end{alignat}
where $i=1,2,3$, and $h_i$ and $f_1$ are functions of a dimensionless variable $x = \frac{U}{U_0}$.
This ansatz is static and possesses SO(9) rotation symmetry, and
if we take $N = \infty$, the metric just becomes the classical solution (\ref{eq:11nearH}).
By solving the equations of motion (\ref{eq:MEOM}), we determine functions $h_i(x)$ and $f_1(x)$.

Calculations are straightforward but complicated, so we use the Mathematica code to explicitly
write down the equations of motion. 
Some of the results are listed in the appendices \ref{sec:app1} and \ref{sec:app2}.
From the output we find that there are five nontrivial equations which are given by
\begin{alignat}{3}
  E_1 &= - 63 x^{34} f_1 - 9 x^{35} f_1' 
  - 49 x^{41} h_1 + 49 x^{34}(1- x^7) h_2 + 23 x^{35}(1- x^7) h_2' + 2 x^{36}(1- x^7) h_2'' \notag
  \\
  &\quad\,
  + 98 x^{41} h_3 + 7 x^{42} h_3' - 63402393600 x^{14} + 70230343680 x^7 + 1062512640 = 0, \label{eq:1}
  \\[0.2cm]
  E_2 &= 63 x^{34} f_1 + 9 x^{35} f_1' + 7 x^{34}(9 - 2 x^{7}) h_1 + 9 x^{35} (1 - x^{7}) h_1' 
  - 112 x^{34} (1 - x^{7}) h_2 \notag
  \\
  &\quad\,
  - 16 x^{35}(1 - x^{7}) h_2' - 98 x^{41} h_3 - 7 x^{42} h_3' - 2159861760 x^7- 5730600960
  = 0, \label{eq:2}
  \\[0.2cm]
  E_3 &= 133 x^{34} f_1 + 35 x^{35} f_1' + 2 x^{36} f_1'' 
  + 28 x^{34} (3 - 10 x^{7}) h_1 + 7 x^{35} (4 - 7 x^{7}) h_1' + 2 x^{36} (1 - x^{7}) h_1'' \notag
  \\
  &\quad\,
  - 7 x^{34} (5 - 26 x^{7}) h_2 - 21 x^{35} (1 - 2 x^{7}) h_2' - 2 x^{36} (1 - x^{7}) h_2'' 
  + 98 x^{41} h_3 + 7 x^{42} h_3' \label{eq:3}
  \\
  &\quad\,
  + 5669637120 x^7 - 8626383360 = 0, \notag
  \\[0.2cm]
  E_4 &= 259 x^{34} f_1 + 53 x^{35} f_1' + 2 x^{36} f_1'' 
  + 147 x^{34}(1 - 3 x^{7}) h_1 + x^{35}(37 -58 x^{7}) h_1' \notag
  \\
  &\quad\,
  + 2 x^{36}(1 - x^{7}) h_1'' 
  + 147 x^{41} h_2 + 21 x^{42} h_2'
  + 294 x^{41} h_3 + 21 x^{42} h_3' \label{eq:4}
  \\
  &\quad\,
  - 63402393600 x^{14} + 133632737280 x^7 - 71292856320 = 0, \notag
  \\[0.2cm]
  E_5 &= 49 x^{34} h_1 + 7 x^{35} h_1' + 49 x^{34} h_2 - x^{35} h_2' - x^{36} h_2'' 
  - 98 x^{34} h_3 - 22 x^{35} h_3' - x^{36} h_3'' \notag
  \\
  &\quad\,
  - 63402393600 x^7 + 70230343680 = 0. \label{eq:5}
\end{alignat}
Here we defined $E_1 = 4U_0^8 \ell_s^4 x^{36}\gamma^{-1} E_{00}$,
$E_2 = 4U_0^8 \ell_s^4 x^{36}\gamma^{-1} E_{11}$, $E_3 = 4U_0^8 \ell_s^4 x^{36}\gamma^{-1} E_{22}$,
$E_4 = 4U_0^8 \ell_s^4 x^{36}\gamma^{-1} E_{1\!0 1\!0}$ and
$E_5 = 4U_0^8 \ell_s^4 x^{\frac{65}{2}}(-1+x^7)^{-\frac{1}{2}}\gamma^{-1} E_{01\!0}$.
Note that the above equations are derived up to the order of $\g$,
and a part of $\g^0$ is zero since the ansatz (\ref{eq:Mansatz}) is a fluctuation
around the classical solution (\ref{eq:11nearH}).

Now we solve these equations to obtain $h_i$ and $f_1$. 
We will see that $h_i$ and $f_1$ are uniquely determined as functions of $x$ by imposing reasonable boundary conditions.
Because calculations below are a bit tedious, the results are summarized in the end of this section.

First let us evaluate the sum of $E_1$ and $E_2$.
\begin{alignat}{3}
  \frac{1}{9x^{28}(x^7-1)} (E_1 + E_2) 
  &= - 7 x^6 h_1 - x^7 h_1' + 7 x^6 h_2 - \frac{7}{9} x^7 h_2' - \frac{2}{9} x^8 h_2'' \notag
  \\
  &\quad\;
  + \frac{518676480}{x^{28}} - \frac{7044710400}{x^{21}} \notag
  \\
  &= \Big( -x^7 h_1  + x^7 h_2 - \frac{2}{9} x^8 h_2'
  + \frac{352235520}{x^{20}} - \frac{19210240}{x^{27}} \Big)' = 0.
\end{alignat}
From this equation $h_1$ is expressed in terms of $h_2$ as
\begin{alignat}{3}
  h_1 = h_2 - \frac{2}{9} x h_2' 
  + \frac{c_1}{x^7} + \frac{352235520}{x^{27}} - \frac{19210240}{x^{34}}, \label{eq:solh1}
\end{alignat}
where $c_1$ is an integral constant. 
Next let us evaluate $E_5$.
\begin{alignat}{3}
  \frac{1}{x^{28}} E_5 &= 49 x^6 h_1 + 7 x^7 h_1' + 49 x^6 h_2 - x^7 h_2' - x^8 h_2'' 
  - 98 x^6 h_3 - 22 x^7 h_3' - x^8 h_3'' \notag
  \\
  &\quad\,
  - \frac{63402393600}{x^{21}} + \frac{70230343680}{x^{28}} \notag
  \\
  &= \Big( 7 x^{7} h_1 + 7 x^{7} h_2 - x^{8} h_2' - 14 x^{7} h_3 - x^{8} h_3' 
  + \frac{3170119680}{x^{20}} -\frac{2601123840}{x^{27}} \Big)' \notag
  \\
  &= \Big( 14 x^{7} h_2 - \frac{23}{9} x^{8} h_2' - 14 x^{7} h_3 - x^{8} h_3' 
  + \frac{5635768320}{x^{20}} -\frac{2735595520}{x^{27}} \Big)' = 0. 
\end{alignat}
In the last line, we removed $h_1$ by using the eq.~(\ref{eq:solh1}).
Thus a linear combination of $h_3$ is expressed in terms of $h_2$ as
\begin{alignat}{3}
  14 x^{7} h_3 + x^{8} h_3' 
  &= 14 x^{7} h_2 - \frac{23}{9} x^{8} h_2' 
   + c_2 + \frac{5635768320}{x^{20}} - \frac{2735595520}{x^{27}}, \label{eq:solh3}
\end{alignat}
where $c_2$ is an integral constant. 
From the eqs.~(\ref{eq:solh1}) and (\ref{eq:solh3}), it is possible to remove
$h_1$ and $h_3$ out of $E_1$, $E_3$ and $E_4$. After some calculations, we obtain
three equations remaining to be solved.
\begin{alignat}{3}
  E_1 &= - 63 x^{34} f_1 - 9 x^{35} f_1' 
  + 49 x^{34} h_2 + x^{35} (23 - 30 x^{7} ) h_2' + 2 x^{36}(1 - x^{7} ) h_2'' \notag
  \\
  &\quad\;
  - 49 c_1 x^{34} + 7 c_2 x^{34} 
  - 41211555840 x^{14} + 52022476800 x^7 + 1062512640 = 0, \label{eq:2-1}
  \\[0.2cm]
  E_3 &= 133 x^{34} f_1 + 35 x^{35} f_1' + 2 x^{36} f_1'' \notag
  \\
  &\quad\,
  + 49 x^{34} h_2 - \frac{7}{9} x^{35} ( 23 - 62 x^{7} ) h_2'
  - \frac{2}{9} x^{36} (32 - 53 x^{7} ) h_2'' 
  - \frac{4}{9} x^{37} ( 1 - x^{7} ) h_2''' \label{eq:2-3}
  \\
  &\quad\,
  - 49 c_1 x^{34} + 7 c_2 x^{34} - 125748080640 x^{14} + 301493283840 x^7 - 37672266240 = 0, \notag
  \\[0.2cm]
  E_4 &= 259 x^{34} f_1 + 53 x^{35} f_1' + 2 x^{36} f_1'' \notag
  \\
  &\quad\,
  + 147 x^{34} h_2
  - \frac{7}{9} x^{35} ( 5 - 26 x^{7} ) h_2'
  - \frac{2}{9} x^{36} ( 32 - 53 x^{7} ) h_2''
  - \frac{4}{9} x^{37} ( 1 - x^{7} ) h_2''' \label{eq:2-4}
  \\
  &\quad\,
  - 147 c_1 x^{34} + 21 c_2 x^{34}
  - 81366405120 x^{14} + 324970168320 x^7 - 95670650880 . \notag
\end{alignat}
Notice, however, that three functions $E_1$, $E_3$ and $E_4$ are not independent because of the identity
\begin{alignat}{3}
  E_4 = \frac{2}{7} x E_1' - 9 E_1 + \frac{16}{7} E_3.
\end{alignat}
This corresponds to the energy conservation, $D_a E^{ab} = 0$.
Thus we only need to solve following two equations.
\begin{alignat}{3}
  -\frac{1}{2}E_1 + \frac{1}{4} (E_3 - E_4) 
  &= -\frac{1}{14}x E_1' + \frac{7}{4} E_1 - \frac{9}{28} E_3 \notag
  \\
  &= - 49 x^{34} h_2 - x^{35} ( 15 - 22 x^{7}) h_2' - x^{36} (1 - x^{7}) h_2'' 
  + 7 ( 7 c_1 - c_2 ) x^{34} \notag
  \\
  &\quad\,
  + 9510359040 x^{14} - 31880459520 x^7 + 13968339840 = 0, \label{eq:3-1}
  \\[0.2cm]
  \frac{1}{2} (E_3 - E_4) 
  &= - \frac{1}{7} x E_1' + \frac{9}{2} E_1 - \frac{9}{14} E_3 \notag
  \\
  &= - 63 x^{34} f_1 - 9 x^{35} f_1' 
  - 49 x^{34} h_2 - 7 x^{35} (1 - 2 x^{7} ) h_2' 
  + 7 ( 7c_1 - c_2) x^{34} \notag
  \\
  &\quad\,
  - 22190837760 x^{14} - 11738442240 x^7 + 28999192320 = 0. \label{eq:3-2}
\end{alignat}
By solving the eq.~(\ref{eq:3-1}), finally we obtain $h_2$ as
\begin{alignat}{3}
  h_2 &= \frac{19160960}{x^{34}} - \frac{58528288}{x^{27}} + \frac{2213568}{13 x^{20}}
  - \frac{1229760}{13x^{13}} \notag
  \\
  &\quad\,
  + c_1 - \frac{c_2}{7} + \frac{2459520}{x^6} + \frac{c_4}{3136 x^7} 
  + 1054080 \Big( 2 - \frac{1}{x^7} \Big) I(x), \label{eq:h2pre}
  \\[0.2cm]
  I(x) &= \frac{c_3}{944455680} + \log (x-1) + \frac{c_4}{6611189760} \log (1-x^{-7}) \notag
  \\[0.2cm]
  &\quad\,
  - \sum_{n=1,3,5} \cos \tfrac{n\pi}{7} \log\big(x^2 + 2 x \cos \tfrac{n\pi}{7} + 1 \big) \notag
  \\[-0.2cm]
  &\quad\,
  - 2 \sum_{n=1,3,5} \sin \tfrac{n\pi}{7} \tan^{-1} \bigg( \frac{x + \cos \tfrac{n\pi}{7} }
  {\sin \tfrac{n\pi}{7}} \bigg), \label{eq:Ipre}
\end{alignat}
where $c_3$ and $c_4$ are integral constants. 
Although the form of $I(x)$ seems to be complicated, its derivative becomes
\begin{alignat}{3}
  I'(x) &= \frac{7}{x^7-1} \Big( 1 + \frac{c_4 \, x^{-1}}{6611189760} \Big). \label{eq:I'pre}
\end{alignat}

So far there are four integral constants, but these will be fixed by appropriate conditions.
In fact it is natural to require that $h_i(1)$ are finite and $h_i(x) \sim \mathcal{O}(x^{-8})$
when $x$ goes to the infinity.
In order to satisfy these conditions, it is necessary to choose
$c_2 = 7 c_1$, $c_3=944455680\pi(\sin\frac{\pi}{7}+\sin\frac{3\pi}{7}+\sin\frac{5\pi}{7})$ 
and $c_4 = -6611189760$.
Inserting these values into the eqs.~(\ref{eq:h2pre}), (\ref{eq:Ipre}) and (\ref{eq:I'pre}), we obtain
\begin{alignat}{3}
  h_2 &= \frac{19160960}{x^{34}}-\frac{58528288}{x^{27}}+\frac{2213568}{13 x^{20}}
  -\frac{1229760}{13x^{13}} \notag
  \\
  &\quad\,
  - \frac{2108160}{x^7} + \frac{2459520}{x^6} 
  + 1054080 \Big( 2 - \frac{1}{x^7} \Big) I(x), \label{eq:h2sol}
  \\[0.2cm]
  I(x) &= \log \frac{x^7(x-1)}{x^7-1} - \sum_{n=1,3,5} \cos \tfrac{n\pi}{7} 
  \log\big(x^2 + 2 x \cos \tfrac{n\pi}{7} + 1 \big) \notag
  \\[-0.1cm]
  &\quad\,
  - 2 \sum_{n=1,3,5} \sin \tfrac{n\pi}{7} \bigg\{
  \tan^{-1} \bigg( \frac{x + \cos \tfrac{n\pi}{7} }
  {\sin \tfrac{n\pi}{7}} \bigg) - \frac{\pi}{2} \bigg\}, \label{eq:I}
\end{alignat}
and
\begin{alignat}{3}
  I'(x) &= \frac{7(1 - x^{-1})}{x^7-1}. \label{eq:I'}
\end{alignat}
Note that the function $I(x)$ behaves as
\begin{alignat}{3}
  I(x) &\sim - \frac{7}{6 x^6} + \frac{1}{x^7} - \frac{7}{13 x^{13}} + \frac{1}{2x^{14}} + \mathcal{O}(x^{-15}),
\end{alignat}
when $x$ goes to the infinity.

Now we remove $h_2$ out of the eq.~(\ref{eq:3-2}), and obtain the differential equation only for $f_1$.
\begin{alignat}{3}
  \frac{1}{18 x^{28}} (E_3 - E_4)
  &= - x^7 f_1' - 7 x^6 f_1 + 819840 I' + 3279360 x^7 (x^7-1) I' \notag
  \\[0.1cm]
  &\quad\,
  + \frac{3624512640}{x^{28}} - \frac{3228113280}{x^{21}} - \frac{5738880}{x^{14}} - \frac{5738880}{x^7} \notag
  \\
  &\quad\,
  + 22955520 x^6 - 22955520 x^7 \notag
  \\[0.2cm]
  &= \Big( - x^7 f_1 + 819840 I - \frac{1208170880}{9x^{27}} \notag
  \\[0.1cm]
  &\qquad
  + \frac{161405664}{x^{20}} + \frac{5738880}{13x^{13}} + \frac{956480}{x^6} \Big)' = 0.
\end{alignat}
Then $f_1$ is solved as
\begin{alignat}{3}
  f_1 &= - \frac{1208170880}{9x^{34}} \!+\! \frac{161405664}{x^{27}} \!+\! \frac{5738880}{13x^{20}}
  \!+\! \frac{956480}{x^{13}} \!+\! \frac{819840}{x^7} I(x).
\end{alignat}
Here the integral constant is set to be zero, because we imposed the boundary condition that
$f_1(x) \sim \mathcal{O}(x^{-8})$ when $x$ goes to the infinity.
From the eq.~(\ref{eq:solh1}), $h_1$ is determined as
\begin{alignat}{3}
  h_1 &= \frac{1302501760}{9x^{34}}-\frac{57462496}{x^{27}}+\frac{12051648}{13
  x^{20}}-\frac{4782400}{13x^{13}} \notag
  \\
  &\quad\,
  - \frac{3747840}{x^7} + \frac{4099200}{x^6} - \frac{1639680 (x-1)}{(x^7-1)} 
  + 117120 \Big( 18 - \frac{23}{x^7} \Big) I(x).
\end{alignat}
The integral constant $c_1$ is chosen to be zero so as to satisfy $h_1(x) \sim \mathcal{O}(x^{-8})$ when 
$x$ goes to the infinity.
Finally from the eq.~(\ref{eq:solh3}), we derive
\begin{alignat}{3}
  0 &= - x^{14} h_3' - 14 x^{13} h_3 + (29514240 x^{13} - 33613440 x^6) I(x) \notag
  \\
  &\quad\,
  + (2693760 x^7 - 5387520 x^{14}) I'(x) + 72145920 x^7 - 67226880 x^6 \notag
  \\
  &\quad\,
  - \frac{7222208000}{9x^{21}} 
  + \frac{777920416}{x^{14}} + \frac{144127872}{13 x^7} - \frac{58072000}{13} \notag
  \\
  &= \Big( - x^{14} h_3 + (2108160 x^{14} - 4801920 x^7) I(x) 
  + 2459520 x^8 - 2108160 x^7 \notag
  \\
  &\qquad
  + \frac{361110400}{9 x^{20}} - \frac{59840032}{x^{13}}
  - \frac{24021312}{13 x^6} - \frac{58072000}{13}x \Big)'.
\end{alignat}
Thus $h_3$ is expressed as
\begin{alignat}{3}
  h_3 &= \frac{361110400}{9x^{34}} - \frac{59840032}{x^{27}}
  - \frac{24021312}{13x^{20}} 
  - \frac{58072000}{13x^{13}} \notag
  \\
  &\quad\, 
  - \frac{2108160}{x^7} + \frac{2459520}{x^6} + 117120 \Big(18 - \frac{41}{x^7} \Big) I(x).
\end{alignat}
The integral constant is set to be zero, since this term can be removed by the general coordinate
transformation on $z$ direction. It corresponds to the gauge transformation on $C_\mu$ in ten dimensions.

Let us summarize the quantum correction to the near horizon geometry of the non-extremal 
M-wave and the black 0-brane.
By solving the eqs.~(\ref{eq:1})--(\ref{eq:5}), 
we obtained the quantum near horizon geometry of the non-extremal M-wave,
\begin{alignat}{3}
  ds_{11}^2 &= \ell_s^4 \Big( - H_1^{-1} F_1 dt^2 + F_1^{-1} U_0^2 dx^2
  + U_0^2 x^2 d\Omega_8^2 + \big( \ell_s^{-4} H_2^{\frac{1}{2}} dz 
  - H_3^{-\frac{1}{2}} dt \big)^2 \Big), \label{eq:Msol}
  \\
  H_i &= \frac{(2\pi)^4 15\pi \lambda}{U_0^7} \Big(\frac{1}{x^7} + \epsilon \frac{\lambda^2}{U_0^6} h_i\Big), \qquad
  F_1 = 1 - \frac{1}{x^7} + \epsilon \frac{\lambda^2}{U_0^6} f_1. \notag
\end{alignat}
In stead of $\gamma$, we introduced dimensionless parameter
\begin{alignat}{3}
  \epsilon = \frac{\gamma}{\lambda^2} = \frac{\pi^6}{2^7 3^2 N^2} \sim \frac{0.835}{N^2},
\end{alignat}
and the functions $h_i$ and $f_1$ are uniquely determined as
\begin{alignat}{3}
  h_1 &= \frac{1302501760}{9x^{34}}-\frac{57462496}{x^{27}}+\frac{12051648}{13
  x^{20}}-\frac{4782400}{13x^{13}}  \notag
  \\
  &\quad\,
  - \frac{3747840}{x^7} + \frac{4099200}{x^6} - \frac{1639680 (x-1)}{(x^7-1)} 
  + 117120 \Big( 18 - \frac{23}{x^7} \Big) I(x), \notag
  \\[0.2cm]
  h_2 &= \frac{19160960}{x^{34}}-\frac{58528288}{x^{27}}+\frac{2213568}{13 x^{20}}
  -\frac{1229760}{13x^{13}} \notag
  \\
  &\quad\,
  - \frac{2108160}{x^7} + \frac{2459520}{x^6} 
  + 1054080 \Big( 2 - \frac{1}{x^7} \Big) I(x), \notag
  \\[0.2cm]
  h_3 &= \frac{361110400}{9x^{34}} - \frac{59840032}{x^{27}}
  - \frac{24021312}{13x^{20}} - \frac{58072000}{13x^{13}} 
  \\
  &\quad\, 
  - \frac{2108160}{x^7} + \frac{2459520}{x^6} 
  + 117120 \Big(18 - \frac{41}{x^7} \Big) I(x), \notag
  \\[0.2cm]
  f_1 &= - \frac{1208170880}{9x^{34}} \!+\! \frac{161405664}{x^{27}} \!+\! \frac{5738880}{13x^{20}}
  \!+\! \frac{956480}{x^{13}} \!+\! \frac{819840}{x^7} I(x). \notag
\end{alignat}
The function $I(x)$ is defined by the eq.~(\ref{eq:I}).
In order to fix the integral constants, we required that $h_i(1)$ are finite and 
$h_i(x), f_1(x) \sim \mathcal{O}(x^{-8})$ when $x$ goes to the infinity.
After the dimensional reduction to ten dimensions, we obtain
\begin{alignat}{3}
  ds_{10}^2 &= \ell_s^2 \Big( - H_1^{-1} H_2^{\frac{1}{2}} F_1 dt^2 + H_2^{\frac{1}{2}} F_1^{-1} U_0^2 dx^2
  + H_2^{\frac{1}{2}} U_0^2 x^2 d\Omega_8^2 \Big), \label{eq:10dimbg}
  \\
  e^\phi &= \ell_s^{-3} H_2^{\frac{3}{4}}, \qquad C = \ell_s^4 H_2^{-\frac{1}{2}} H_3^{-\frac{1}{2}} dt.
  \notag
\end{alignat}
This represents the quantum near horizon geometry of the non-extremal black 0-brane.

\section{Thermodynamics of Quantum Near Horizon Geometry of Black 0-Brane}\label{sec:thermo}

Since the quantum near horizon geometry of the non-extremal black 0-brane is derived in the previous section,
it is interesting to evaluate its thermodynamics.
In this section, we estimate the entropy and the internal energy of the quantum near horizon geometry of
the non-extremal black 0-brane by using Wald's formula~\cite{Wa,IW}. 
These quantities are quite important when we test the gauge/gravity duality.

In the following, quantities are calculated up to $\mathcal{O}(\epsilon^2)$.
First of all, let us examine the location of the horizon $x_\text{H}$. 
This is defined by $F_1(x_\text{H}) = 0$ and becomes
\begin{alignat}{3}
  x_\text{H} = 1 - \epsilon \frac{f_1(1)}{7} \tilde{U}_0^{-6},
\end{alignat}
where $\tilde{U}_0 \equiv U_0/\lambda^{\frac{1}{3}}$ is a dimensionless parameter.
Temperature of the black 0-brane is derived by the usual prescription.
We consider the Euclidean geometry by changing time coordinate as $t=-i\tau$
and require the smoothness of the geometry at the horizon.
This fixes the periodicity of $\tau$ direction and its inverse gives the temperature of the non-extremal black 0-brane.
Then the dimensionless temperature $\tilde{T} = T/\lambda^{\frac{1}{3}}$ of the black 0-brane is evaluated as
\begin{alignat}{3}
  \tilde{T} &= \frac{1}{4\pi} U_0^{-1} H_1^{-\frac{1}{2}} F_1' \Big|_{x_\text{H}} \Big/ \lambda^{\frac{1}{3}}
  = a_1 \tilde{U}_0^\frac{5}{2} \big( 1 + \epsilon a_2 \tilde{U}_0^{-6} \big), \label{eq:T}
\end{alignat}
where $a_1$ and $a_2$ are numerical constants given by
\begin{alignat}{3}
  a_1 &= \frac{7}{16\pi^3\sqrt{15\pi}} \sim 0.00206, \notag
  \\
  a_2 &= \frac{9}{14} f_1(1) + \frac{1}{7} f'_1(1) - \frac{1}{2} h_1(1) \sim 937000. 
\end{alignat}
Inversely solving the eq.~(\ref{eq:T}), the dimensionless parameter $\tilde{U}_0$ is written
in terms of the temperature $\tilde{T}$ as
\begin{alignat}{3}
  \tilde{U}_0 &= a_1^{-\frac{2}{5}} \tilde{T}^\frac{2}{5}
  \Big( 1 - \epsilon \frac{2}{5} a_1^{\frac{12}{5}} a_2 \tilde{T}^{-\frac{12}{5}} \Big), \label{eq:tildeU0}
\end{alignat}
By using this replacement, it is always possible to express physical quantities
as functions of $\tilde{T}$.

Next we derive the entropy of the quantum near horizon geometry of the non-extremal black 0-brane.
In practice, we consider the quantum near horizon geometry of the non-extremal M-wave
because of its simple expression.
Since the effective action (\ref{eq:Maction}) includes higher derivative terms, we should employ Wald's entropy formula
which ensures the first law of the black hole thermodynamics.
The Wald's entropy formula is given by
\begin{alignat}{3}
  S &= -2\pi \int_\text{H} d\Omega_8 dz \sqrt{h}
  \frac{\partial S_{11}}{\partial R_{\mu\nu\rho\sigma}} N_{\mu\nu} N_{\rho\sigma}, \label{eq:Wald}
\end{alignat}
where $\sqrt{h} = (\ell_s^2U_0x)^8\ell_s^{-2}H_2^{1/2}$ 
is the volume factor at the horizon and $N_{\mu\nu}$ is an antisymmetric tensor binormal to the horizon. 
The binormal tensor satisfies $N_{\mu\nu}N^{\mu\nu} = -2$
and nonzero component is only $N_{tx} = - \ell_s^4 U_0 H_1^{-1/2}$.
The effective action is given by the eq.~(\ref{eq:Maction}), and
in the formula the variation of the action is evaluated as if the Riemann tensor 
is an independent variable, that is, 
\begin{alignat}{3}
  \frac{\partial S_{11}}{\partial R_{\mu\nu\rho\sigma}} &=
  \frac{1}{2\kappa_{11}^2} \big( g^{\mu[\rho}g^{\sigma]\nu} 
  + \gamma \ell_s^{12} X^{\mu\nu\rho\sigma} \big). \label{eq:varS}
\end{alignat}
Now we are ready to evaluate the entropy of the quantum near horizon geometry of the non-extremal M-wave.
Some useful results are collected in the appendix \ref{sec:app2}.
By using these, the entropy is evaluated as
\begin{alignat}{3}
  S &= \frac{4\pi}{2\kappa_{11}^2} \int_\text{H} d\Omega_8 dz \sqrt{h}
  \Big( 1 - \frac{1}{2} \gamma \ell_s^{12} X^{\mu\nu\rho\sigma} 
  N_{\mu\nu} N_{\rho\sigma} \Big) \notag
  \\
  &= \frac{4\pi}{2\kappa_{11}^2} \int_\text{H} d\Omega_8 dz \sqrt{h}
  \big( 1 - 2 \gamma \ell_s^{20} U_0^2 H_1^{-1} X^{txtx} \big) \notag
  \\
  &= \frac{4\pi}{2\kappa_{11}^2} \int_\text{H} d\Omega_8 dz \sqrt{h}
  \Big( 1 + 40642560 \epsilon \frac{1}{\tilde{U}_0^6 x_\text{H}^{20}} \Big) \notag
  \\
  &= \frac{4}{49} a_1 N^2 \tilde{U}_0^\frac{9}{2}
  \Big\{ 1 + \epsilon \Big( - \frac{9}{14} f_1(1) + \frac{1}{2} h_2(1) + 40642560 \Big) 
  \tilde{U}_0^{-6} \Big\} \notag
  \\
  &= \frac{4}{49} a_1^{-\frac{4}{5}} N^2 \tilde{T}^\frac{9}{5}
  \Big\{ 1 + \epsilon a_1^{\frac{12}{5}}\Big( - \frac{9}{5} f_1(1) - \frac{9}{35} f'_1(1)
  + \frac{9}{10} h_1(1) + \frac{1}{2} h_2(1) + 40642560 \Big) 
  \tilde{T}^{-\frac{12}{5}} \Big\} \notag
  \\[0.1cm]
  &= a_3 N^2 \tilde{T}^\frac{9}{5}
  \big( 1 + \epsilon \, a_4 \tilde{T}^{-\frac{12}{5}} \big), \label{eq:entropy}
\end{alignat}
where numerical constants $a_3$ and $a_4$ are defined as
\begin{alignat}{3}
  a_3 &= \frac{4}{49} a_1^{-\frac{4}{5}} 
  = 2^{\frac{26}{5}} 15^{\frac{2}{5}} 7^{-\frac{14}{5}} \pi^\frac{14}{5} \sim 11.5, \notag
  \\
  a_4 &= a_1^{\frac{12}{5}} \Big( - \frac{9}{5}f_1(1) - \frac{9}{35}f'_1(1) 
  + \frac{9}{10}h_1(1) + \frac{1}{2}h_2(1) + 40642560 \Big) \sim 0.400.
\end{alignat}
So far we have obtained the entropy for the M-wave.
Because of the duality between type IIA string theory and M-theory, 
this is equivalent to that of the black 0-brane.

Finally let us derive the internal energy of the quantum near horizon geometry of the non-extremal black 0-brane.
Wald's entropy formula is constructed so as to satisfy the thermodynamic laws of black holes.
Then by integrating $d\tilde{E} = \tilde{T} dS$, it is possible to obtain the dimensionless 
energy $\tilde{E} = E/\lambda^{\frac{1}{3}}$ as
\begin{alignat}{3}
  \frac{\tilde{E}}{N^2} &= \frac{9}{14} a_3 \tilde{T}^\frac{14}{5}
  - \epsilon \frac{3}{2} a_3 a_4 \tilde{T}^{\frac{2}{5}}
  \sim 7.41 \tilde{T}^\frac{14}{5} - \frac{5.77}{N^2} \tilde{T}^{\frac{2}{5}}. \label{eq:energy}
\end{alignat}
This result includes the quantum gravity effect, and it gives quite nontrivial test of the gauge/gravity
duality if we can evaluate the internal energy from the dual gauge theory.
In fact it is possible by employing the Monte Carlo simulation and the result strongly concludes
that the duality holds at this order~\cite{HHIN}.

The specific heat is evaluated as
\begin{alignat}{3}
  \frac{1}{N^2} \frac{d\tilde{E}}{d\tilde{T}} &= 
  \frac{9}{5} a_3 \tilde{T}^\frac{9}{5} - \epsilon \frac{3}{5} a_3 a_4 \tilde{T}^{-\frac{3}{5}}. \label{eq:heat}
\end{alignat}
Notice that the specific heat becomes negative in the region where $\tilde{T} <
(\epsilon a_4/3)^{5/12} \sim 0.4 N^{-5/6}$. In this region the non-extremal black 0-brane
behaves like Schwarzschild black hole and will be unstable. When $N=\infty$ the instability will be suppressed.
This result is also verified from the Monte Carlo simulation of the dual gauge theory~\cite{HHIN}.

\section{D0-brane Probe}\label{sec:probe}

In this section, we probe the quantum near horizon geometry of the non-extremal black 0-brane 
(\ref{eq:10dimbg}) via a D0-brane. Form the analysis it is possible to study how
the test D0-brane is affected by the background field.

The bosonic part of the D0-brane action consists of the Born-Infeld action
and the Chern-Simons one. 
Here we neglect an excitation of the gauge field on the D0-brane, so
the Born-Infeld action is simply given by the pull-back of the metric.
We also assume that the D0-brane moves only along the radial direction.
Then the probe D0-brane action in the background of (\ref{eq:10dimbg}) is written as
\begin{alignat}{3}
  S_{\text{D}0} &= - T_0 \int dt e^{-\phi} \sqrt{- g_{\mu\nu} \frac{dx^\mu}{dt}\frac{dx^\nu}{dt}}
  + T_0 \int C \notag
  \\
  &= - T_0 \ell_s^4 \int dt H_2^{-\frac{1}{2}} \sqrt{H_1^{-1} F_1 - F_1^{-1} U_0^2 \dot{x}^2}
  + T_0 \ell_s^4 \int dt H_2^{-\frac{1}{2}} H_3^{-\frac{1}{2}}.
\end{alignat}
The momentum conjugate to $x$ is evaluated as
\begin{alignat}{3}
  p = T_0 \ell_s^4 H_2^{-\frac{1}{2}} \frac{F_1^{-1} U_0^2\dot{x}}{\sqrt{H_1^{-1} F_1 - F_1^{-1} U_0^2 \dot{x}^2}},
\end{alignat}
and the energy of the probe D0-brane is given by
\begin{alignat}{3}
  E_{\text{D}0} &= p \dot{x} + T_0 \ell_s^4 H_2^{-\frac{1}{2}} \sqrt{H_1^{-1} F_1 - F_1^{-1} U_0^2 \dot{x}^2}
  - T_0 \ell_s^4 H_2^{-\frac{1}{2}} H_3^{-\frac{1}{2}} \notag
  \\
  &= T_0 \ell_s^4 H_2^{-\frac{1}{2}} \frac{H_1^{-1} F_1}{\sqrt{H_1^{-1} F_1 - F_1^{-1} U_0^2 \dot{x}^2}}
  - T_0 \ell_s^4 H_2^{-\frac{1}{2}} H_3^{-\frac{1}{2}} \notag
  \\
  &= T_0 \ell_s^4 H_1^{-\frac{1}{2}} H_2^{-\frac{1}{2}} F_1^{\frac{1}{2}} 
  \sqrt{1 + \Bigg( \frac{p F_1^{\frac{1}{2}} H_2^{\frac{1}{2}}}{T_0 \ell_s^4 U_0} \Bigg)^2}
  - T_0 \ell_s^4 H_2^{-\frac{1}{2}} H_3^{-\frac{1}{2}} \notag
  \\
  &\sim \frac{1}{2} H_1^{-\frac{1}{2}} H_2^{\frac{1}{2}} F_1^{\frac{3}{2}} 
  \frac{p^2}{T_0 \ell_s^4 U_0^2}
  + T_0 \ell_s^4 \Big( H_1^{-\frac{1}{2}} H_2^{-\frac{1}{2}} F_1^{\frac{1}{2}} 
  - H_2^{-\frac{1}{2}} H_3^{-\frac{1}{2}} \Big).
\end{alignat}
In the final line we took the non-relativistic limit.
From this we see that the potential energy for the probe D0-brane is expressed as
\begin{alignat}{3}
  V_{\text{D}0} &= T_0 \ell_s^4 \Big( H_1^{-\frac{1}{2}} H_2^{-\frac{1}{2}} F_1^{\frac{1}{2}} 
  - H_2^{-\frac{1}{2}} H_3^{-\frac{1}{2}} \Big) .
\end{alignat}
The first term corresponds to the gravitational attractive force and
the second one does to the R-R repulsive force.

\begin{figure}[htbp]
\begin{center}
\begin{picture}(300,220)
\put(290,5){$x$}
\includegraphics[width=10cm]{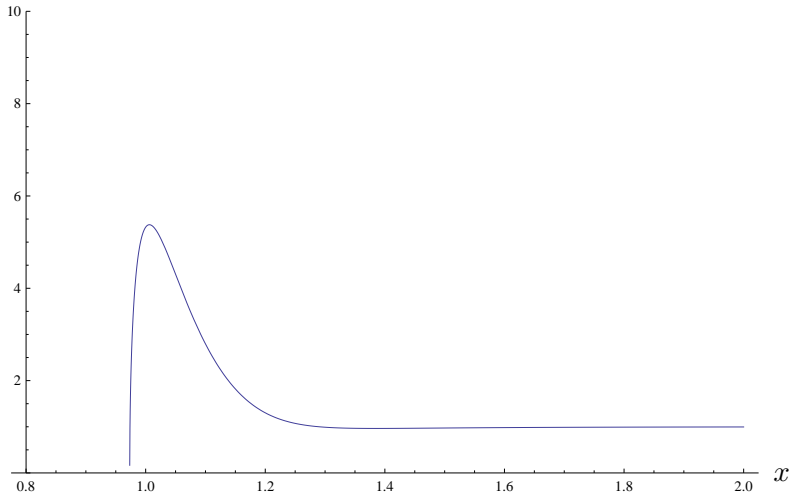}
\end{picture}
\caption{The function $\sqrt{F_1(x)}$ with $F_1(x) = 1-1/x^7+0.000001 f_1(x).$
} \label{fig:F_1}
\end{center}
\end{figure}

When we take $N=\infty$, the potential energy becomes 
$V_{\text{D}0} = T_0 \ell_s^4 H^{-1} (\sqrt{F}-1)$.
The part $(\sqrt{F}-1)$ shows that the gravitational attractive force overcomes the R-R repulsive force.
Similarly, when $N$ is finite, we regard $\sqrt{F_1}$ as the gravitational attractive force
to the probe D0-brane. The function of $\sqrt{F_1}$ is plotted in fig.~{\ref{fig:F_1}}.
From this we see that the gravitational force becomes repulsive near the horizon $x_\text{H}$.

\section{Validity of the Analyses on Quantum Near Horizon Geometry}\label{sec:valid}

Our analyses so far are based on the effective action (\ref{eq:Maction}), which becomes
the 1-loop effective action of the type IIA superstring theory after the dimensional reduction.
Since the superstring theory is defined by the perturbative expansions of $\alpha'$ and $g_s$,
terms with higher powers of these parameters also contribute to the effective action.
Then our results are valid when the 1-loop effect becomes dominant compared to other
stringy or loop effects. In this section we clarify the valid parameter region of our analyses.

First let us consider the validity of the type IIA supergravity approximation.
From eq.~(\ref{eq:nearH}), the curvature radius $\rho$
and the effective string coupling $g_s e^{\phi}$ at the event horizon $U=U_0$ are evaluated as
\begin{alignat}{3}
  \frac{\alpha'}{\rho^2} \sim \tilde{U}_0^{\frac{3}{2}} \sim \tilde{T}^{\frac{3}{5}}, \qquad 
  g_s e^{\phi} \sim \frac{\tilde{U}_0^{-\frac{21}{4}}}{N} \sim \frac{\tilde{T}^{-\frac{21}{10}}}{N}.
  \label{eq:dimana}
\end{alignat}
Here we used the relation eq.~(\ref{eq:tildeU0}) by setting $\epsilon=0$.
Then the supergravity approximation is valid when the string length $\sqrt{\alpha'}$ is 
quite small compared to the curvature radius $\rho$
and the effective string coupling $g_s e^{\phi}$ is also quite small, i.e., 
$\tilde{T} \sim 0$ and $N \sim \infty$.

\begin{figure}[tbp]
\begin{center}
\begin{picture}(250,250)
\put(235,5){$\tilde{T}$}
\put(10,225){$N$}
\includegraphics[width=8cm]{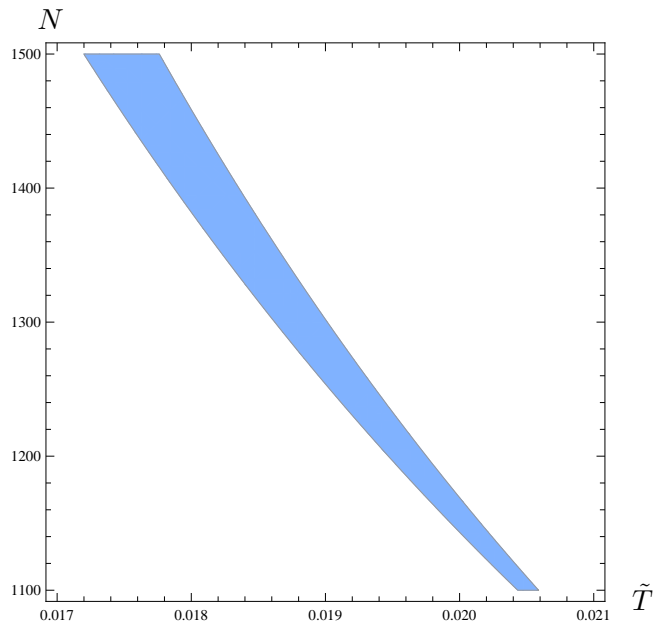}
\end{picture}
\caption{Region of $\frac{1}{\tilde{T}^{9/5}} \leq N \leq \frac{1}{\sqrt{10}\tilde{T}^{21/10}}$.
} \label{fig:region}
\end{center}
\end{figure}

Now we consider the validity of our 1-loop analyses. 
From the effective action (\ref{eq:Maction}), we derived the internal energy~(\ref{eq:energy}) of 
the black 0-brane. 
However, if we include other higher derivative terms in the effective action, 
the Lagrangian is expected to be
\begin{alignat}{3}
  \mathcal{L}
  &\sim R + \big( \alpha'^3 R^4 + \alpha'^5 \partial^4 R^4 + \cdots \big)
  + g_s^2 \big( \alpha'^3 R^4 + \alpha'^6 \partial^6 R^4 + \cdots \big) \notag
  \\
  &\qquad\quad
  + g_s^4 \big( \alpha'^5 \partial^4 R^4 + \cdots \big) + \cdots 
  + g_s^{2n} \big( \alpha'^{3+n} \partial^{2n} R^4 + \cdots \big) + \cdots, \label{eq:actexp}
\end{alignat}
where $R$ is the abbreviation of the Riemann tensor.
The existence of these terms can be found in ref.~\cite{GRV}.
By following the calculation of eq.~(\ref{eq:entropy}) and using the dimensional analyses (\ref{eq:dimana}), 
the internal energy will be modified as
\begin{alignat}{3}
  \frac{\tilde{E}}{N^2} 
  &\sim 7.41 \tilde{T}^\frac{14}{5} \Big\{ 1 + \Big( \tilde{T}^{\frac{9}{5}} + \tilde{T}^{3} + \cdots \Big)
  + \frac{1}{N^2} \Big( - \frac{0.779}{\tilde{T}^{\frac{12}{5}}} 
  + \frac{1}{\tilde{T}^{\frac{3}{5}}} + \cdots \Big) \notag
  \\
  &\qquad\qquad\quad
  + \frac{1}{N^4} \Big( \frac{c_2}{\tilde{T}^{\frac{27}{5}}} + \cdots \Big) + \cdots 
  + \frac{1}{N^{2n}} \Big( \frac{1}{\tilde{T}^{\frac{18}{5}n-\frac{9}{5}}} + \cdots \Big) 
  + \cdots \Big\} . \label{eq:energy2}
\end{alignat}
$\tilde{T}^\frac{9}{5}$ and $\tilde{T}^3$ come from $\alpha'^3$ and $\alpha'^5$ terms at tree level, and
$\frac{1}{N^{2n}}$ correspond to $n$-loop amplitudes.
Numerical constants are assumed to be $\mathcal{O}(1)$ and at least this is true for the 1-loop result.
The coefficient $c_2$ at 2-loop will be discussed later.
From the above estimation, the 1-loop contribution of $-0.779/(N^2\tilde{T}^{\frac{12}{5}})$ 
becomes subleading when following conditions are satisfied,
\begin{alignat}{3}
  \frac{\tilde{T}^{\frac{9}{5}}}{1/N^2 \tilde{T}^{\frac{12}{5}}} 
  = N^2 \tilde{T}^{\frac{21}{5}} \leq s, \qquad
  \tilde{T}^{\frac{9}{5}} \leq s, \qquad
  \frac{1/N^{2n} \tilde{T}^{\frac{18n}{5}-\frac{9}{5}}}{1/N^2 \tilde{T}^{\frac{12}{5}}} 
  = \frac{N^2 \tilde{T}^{\frac{21}{5}}}{N^{2n} \tilde{T}^{\frac{18n}{5}}} \leq s,
\end{alignat}
where $s < 1$. 
The above inequalities are equivalent to
\begin{alignat}{3}
  \frac{1}{\tilde{T}^{\frac{9}{5}}} \leq N \leq \frac{\sqrt{s}}{\tilde{T}^{\frac{21}{10}}}. 
  \label{eq:valid}
\end{alignat}
Thus our analyses are estimated to be valid in this parameter region. 

The case of $s=0.1$ is drawn in fig.~\ref{fig:region} and
for example $(\tilde{T},N) = (0.02,1150)$ is located inside the region.
Then from eq.~(\ref{eq:tildeU0}), we obtain $U_0 = 2.48$ and $F_1(x) = 1 - 1/x^7 + 0.00357 f_1(x)$. 
This shows that the quantum effect becomes important near the event horizon. (see fig.~\ref{fig:F_1}.)

Notice that the validity of the parameter region obtained in eq.~(\ref{eq:valid}) is roughly estimated.
In order to know more precise one, we should determine the coefficient $c_2$ at 2-loop and others.
Although it is beyond the scope of our paper, if we suppose $c_2 \sim 0.005$ and 
other coefficients $c_n\,(n>2)$ are negligible, 
the lower bound in eq.~(\ref{eq:valid}) is enlarged like $0.0801/\sqrt{s}\tilde{T}^{3/2} \leq N$.
This overlaps with the region $N < 0.334/\tilde{T}^{6/5}$ 
where the specific heat (\ref{eq:heat}) becomes negative, when we choose $s\sim 1$.
For example, $(\tilde{T},N) = (0.02,30)$ is inside the overlap region. 
On the other hand, if we suppose $c_2 \sim 1$, the parameter region (\ref{eq:valid}) does not overlap with
that of negative specific heat. However, as the temperature decreases from the region (\ref{eq:valid}) with $N$ fixed,
the 2-loop term dominates the internal energy. Then if $c_2$ is negative, the internal energy takes
large negative value becuase of the negative power in $c_2 \tilde{T}^{-13/5}$.
Thus we expect $c_2$ is positive, and again the specific heat becomes negative.
As a reference we mention that the numerical simulation suggests that the 2-loop coeficient
becomes $c_2=+0.00459$~\cite{HHIN}.

\section{Conclusion and Discussion}

In this paper we studied quantum nature of the bunch of D0-branes in the type IIA superstring theory.
In the classical limit, it is well described by the non-extremal black 0-brane in the type IIA supergravity.
The quantum correction to the non-extremal black 0-brane is investigated after taking the near horizon limit.

In order to manage the quantum effect of the gravity, 
we uplifted the near horizon geometry of the non-extremal black 0-brane 
into that of the M-wave solution in the eleven dimensional supergravity. 
These two are equivalent via the duality between the type IIA superstring theory and the M-theory,
but the latter is purely geometrical and calculations become rather simple.
The geometrical part of the effective action for the M-theory (\ref{eq:Maction}) is derived 
so as to be consistent with the 1-loop amplitudes in the type IIA superstring theory.
And the quantum correction to the M-wave solution is 
taken into account by explicitly solving the equations of motion (\ref{eq:MEOM}).
The solution is uniquely determined and its explicit form is given by the eq.~(\ref{eq:10dimbg}).
It is interesting to note that a probe D0-brane moving in this
background would feel repulsive force near the horizon.
It means that the solution includes the back-reaction of the Hawking radiation.

We also investigated the thermodynamic property of the quantum near horizon geometry of the non-extremal black 0-brane.
Since the effective action contains higher derivative terms,
we examined the thermodynamic property of the black 0-brane by employing Wald's formula.
The entropy and the internal energy of the black 0-brane are evaluated up to $1/N^2$.
The quantum correction to the internal energy becomes important when $N$ is small.
In ref.~\cite{HHIN}, the internal energy is also calculated from the dual thermal gauge theory 
by using the Monte Carlo simulation, and it agrees with the eq.~(\ref{eq:energy}) very well.
This gives a strong evidence for the gauge/gravity duality at the level of quantum gravity.


Finally we give an important remark on the effective action for the M-theory. 
It contains higher derivative terms, but these cannot be determined uniquely because of the field redefinitions.
In the appendices we have considered all possible higher derivative terms
and shown that the ambiguities of the effective action have nothing to do with 
the thermodynamic properties of near horizon geometry of the non-extremal black 0-brane. 

As a future work, it is important to derive quantum geometry of the non-extremal black 0-brane and
obtain the solution (\ref{eq:10dimbg}) by taking the near horizon limit.
The result will be reported elsewhere, but it is really possible.
It is also interesting to examine quantum correction to the black 6-brane, which is also described by
purely geometrical object, called Kaluza-Klein monopole, in the eleven dimensional supergravity.
To find connections of our results to the other approaches to the field theory on the D0-branes is important 
as well~\cite{LSWY,MSWW}. Other approaches to probe curvature corrections by the black brane will also be 
related to our results in section \ref{sec:probe}~\cite{CF,Cz}.
Since now we capture the quantum nature of the near horizon geometry of the black 0-brane,
it is interesting to consider a recent proposal to resolve the information paradox on 
the black hole~\cite{Ma,BPZ,AMPS}.

\section*{Acknowledgement}

The author would like to thank Masanori Hanada, Goro Ishiki and Jun Nishimura for discussions and
collaborations. He would also like to thank Masafumi Fukuma, Hideki Ishihara, Hikaru Kawai and Yukinori Yasui 
for discussions. This work was partially supported by the Ministry of Education, Science, 
Sports and Culture, Grant-in-Aid for Young Scientists (B) 19740141,
2007 and 24740140, 2012.

\newpage
\appendix

\section{Calculations of Ricci Tensor and Scalar Curvature} \label{sec:app1}

By using the ansatz (\ref{eq:Mansatz}) for the metric,
each component of the Ricci tensor up to the linear order of $\gamma$ is calculated as
\begin{alignat}{3}
  R_{00} &= \frac{\gamma}{4 U_0^8 x^2 \ell_s^4} 
  \big\{ 98 f_1 + 30 x f_1' + 2 x^2 f_1'' 
  + 49 (2 - 7 x^7) h_1 + 3x (10 - 17 x^7) h_1' + 2x^2 (1 - x^7) h_1'' \notag
  \\
  &\quad\,
  + 147 x^7 h_2 + 21 x^8 h_2' + 196 x^7 h_3 + 14 x^8 h_3' \big\}, \notag
  \\[0.1cm]
  R_{11} &= \frac{\gamma}{4 U_0^8 x^2 \ell_s^4} 
  \big\{ - 98 f_1 - 30 x f_1' - 2 x^2 f_1'' 
  - 35 (1 - 8 x^7)h_1 - 21 x (1 - 2 x^7) h_1' - 2 x^2 (1 - x^7) h_1'' \notag
  \\
  &\quad\,
  - 7 (9 + 12 x^7) h_2 + 7x (1 - 4 x^7) h_2' + 2x^2 (1-x^7) h_2'' 
  - 196 x^7 h_3 - 14 x^8 h_3' \big\}, 
  \\[0.1cm]
  R_{\bar{a}\bar{a}} &= \frac{\gamma}{2 U_0^8 x^2 \ell_s^4} 
  \big\{ - 14 f_1 - 2 x f_1' - 7 (1-x^7)h_1 - x (1-x^7) h_1' + 7(1-x^7) h_2 + x(1-x^7) h_2' \big\}, \notag
  \\[0.1cm]
  R_{\natural\natural} &= \frac{\gamma}{4 U_0^8 x^2 \ell_s^4} 
  \big\{ 98 f_1 + 14 x f_1' + 49 (1 - 3x^7) h_1 + 7x (1-x^7) h_1' \notag
  \\
  &\quad\,
  + 49(1-x^7) h_2 + 23x (1-x^7) h_2' + 2x^2 (1-x^7) h_2'' + 196 x^7 h_3 + 14 x^8 h_3' \big\}, \notag
  \\[0.1cm]
  R_{0\natural} &= \frac{\gamma\, x^{3/2}\sqrt{x^7-1}}{4 U_0^8 \ell_s^4} 
  \big\{ 49 h_1 + 7 x h_1' + 49 h_2 - x h_2' - x^2 h_2'' - 98 h_3 - 22 x h_3' - x^2 h_3'' \big\}. \notag
\end{alignat}
Here we used $\natural$ instead of $10$ and $\bar{a}=2,\cdots,9$.
Ricci scalar up to the linear order of $\gamma$ becomes like
\begin{alignat}{3}
  R &= \frac{\gamma}{2 U_0^8 x^2 \ell_s^4} 
  \big\{ - 161 f_1 - 39 x f_1' - 2 x^2 f_1'' 
  - 98 (1 - 3 x^7) h_1 - 3x (10 - 17 x^7) h_1' - 2x^2 (1 - x^7) h_1'' \notag
  \\
  &\quad\,
  + 49 (1 - 4 x^7) h_2 + x(23 - 44 x^7) h_2' + 2x^2 (1-x^7) h_2'' 
  - 98 x^7 h_3 - 7 x^8 h_3' \big\}.
\end{alignat}

\section{Calculations of Higher Derivative Terms} \label{sec:app2}

In this appendix we summarize the values of higher derivative terms appeared in the eq.~(\ref{eq:MEOM}).
Note that we only need to evaluate these terms by using the ansatz (\ref{eq:Mansatz}) with $\gamma=0$,
because the equations of motion are solved up to the linear order of $\gamma$.
First of all, each component of $R_{abcd}$ is calculated as
\begin{alignat}{3}
  R_{0101} &= - \frac{28}{U_0^2 x^2 \ell_s^4}, &\qquad&
  R_{0\bar{a}0\bar{a}} = \frac{7}{2 U_0^2 x^2 \ell_s^4}, \notag
  \\
  R_{011\natural} &= \frac{28 \sqrt{x^7-1}}{U_0^2 x^{\frac{11}{2}} \ell_s^4}, &\qquad&
  R_{0\bar{a}\bar{a}\natural} = - \frac{7 \sqrt{x^7-1}}{2 U_0^2 x^{\frac{11}{2}} \ell_s^4}, \notag
  \\
  R_{1\natural1\natural} &= -\frac{28(x^7-1)}{U_0^2 x^9 \ell_s^4}, &\qquad&
  R_{1\bar{a}1\bar{a}} = - \frac{7}{2 U_0^2 x^9 \ell_s^4}, \label{eq:Riemann}
  \\
  R_{\bar{a}\natural \bar{a}\natural} &= \frac{7(x^7-1)}{2U_0^2 x^9 \ell_s^4}, &\qquad&
  R_{\bar{a}\bar{b}\bar{a}\bar{b}} = \frac{1}{U_0^2 x^9 \ell_s^4}. \notag
\end{alignat}
We used $\natural$ instead of $10$ and $\bar{a},\bar{b}=2,\cdots,9$.
The scalar curvature and each component of the Ricci tensor become zero,
and each component of $X_{abcd}$ in the eq.~(\ref{eq:X}) is evaluated as
\begin{alignat}{3}
  X_{0101} &= - \frac{20321280}{U_0^6 x^{20} \ell_s^{12}}, &\qquad&
  X_{0\bar{a}0\bar{a}} = - \frac{1270080}{U_0^6 x^{20} \ell_s^{12}}, \notag
  \\
  X_{011\natural} &= \frac{20321280 \sqrt{x^7-1}}{U_0^6 x^{\frac{47}{2}} \ell_s^{12}}, &\qquad&
  X_{0\bar{a}\bar{a}\natural} = \frac{1270080 \sqrt{x^7-1}}{U_0^6 x^{\frac{47}{2}} \ell_s^{12}}, \notag
  \\
  X_{1\natural1\natural} &= -\frac{20321280(x^7-1)}{U_0^6 x^{27} \ell_s^{12}}, &\qquad&
  X_{1\bar{a}1\bar{a}} = \frac{1270080}{U_0^6 x^{27} \ell_s^{12}}, 
  \\
  X_{\bar{a}\natural \bar{a}\natural} &= -\frac{1270080(x^7-1)}{U_0^6 x^{27} \ell_s^{12}}, &\qquad&
  X_{\bar{a}\bar{b}\bar{a}\bar{b}} = \frac{1192320}{U_0^6 x^{27} \ell_s^{12}}. \notag
\end{alignat}
By using these results we are ready to calculate higher derivative terms in the eq.~(\ref{eq:MEOM}).
The $R^4$ terms are calculated as
\begin{alignat}{3}
  t_8 t_8 R^4 - \frac{1}{4!} \e_{11} \e_{11} R^4 = \frac{531256320}{U_0^8 x^{36} \ell_s^{16}}.
\end{alignat}
The $RX$ terms become
\begin{alignat}{3}
  &R_{abc0} X^{abc}{}_0 = - \frac{1066867200}{U_0^8 x^{29} \ell_s^{16}}, \quad
  R_{abc1} X^{abc}{}_1 = \frac{1066867200}{U_0^8 x^{36} \ell_s^{16}}, \notag
  \\
  &R_{abc\natural} X^{abc}{}_\natural = - \frac{1066867200(x^7-1)}{U_0^8 x^{36} \ell_s^{16}}, \quad
  R_{abc\bar{a}} X^{abc}{}_{\bar{b}} = - \frac{1088640}{U_0^8 x^{36} \ell_s^{16}} \delta_{\bar{a}\bar{b}}, 
  \\
  &R_{abc0} X^{abc}{}_\natural = R_{abc\natural} X^{abc}{}_{0} = 
  - \frac{1066867200\sqrt{x^7-1}}{U_0^8 x^{\frac{65}{2}} \ell_s^{16}}, \notag
\end{alignat}
and the $DDX$ terms are evaluated as
\begin{alignat}{3}
  &D_{(a} D_{b)} X^a{}_{00}{}^b = \frac{198132480 (-47 + 40 x^7)}{U_0^8 x^{29} \ell_s^{16}}, \quad
  D_{(a} D_{b)} X^a{}_{11}{}^b = \frac{2177280 (513 + 124 x^7)}{U_0^8 x^{36} \ell_s^{16}}, \notag
  \\
  &D_{(a} D_{b)} X^a{}_{\natural\natural}{}^b = 
  \frac{198132480 (47 - 87 x^7 + 40 x^{14})}{U_0^8 x^{36} \ell_s^{16}}, \quad
  D_{(a} D_{b)} X^a{}_{\bar{a}\bar{b}}{}^b = \frac{236234880 (4 - 3 x^7)}{U_0^8 x^{36} \ell_s^{16}} 
  \delta_{\bar{a}\bar{b}}, \notag
  \\
  &D_{(a} D_{b)} X^a{}_{0\natural}{}^b = D_{(a} D_{b)} X^a{}_{\natural 0}{}^b = 
  \frac{198132480 (-47 + 40 x^7)\sqrt{x^7-1}}{U_0^8 x^{\frac{65}{2}} \ell_s^{16}}.
\end{alignat}
By inserting these results into the eq.~(\ref{eq:MEOM}), we obtain the eqs.~(\ref{eq:1})--(\ref{eq:5}).

\section{Generic $R^4$ Terms, Equations of Motion and Solution}
\label{sec:R4}

In this appendix, we classify independent R4 terms which consist of four products of
the Riemann tensor, the Ricci tensor or the scalar curvature. The R4 terms which include
the Ricci tensor or the scalar curvature cannot be determined from the scattering amplitudes
in the type IIA superstring theory. So in general the effective action and equations of motion
are affected by these ambiguities.

First let us review the R4 terms which only consist of the Riemann tensor.
Since there are 16 indices, we have 8 pairs to be contracted.
Naively it seems that there are so many possible patterns.
However, carefully using properties of the Riemann tensor, such as $R_{abcd} = - R_{bcad} - R_{cabd}$,
it is possible to show that there are only 7 independent terms.
\begin{alignat}{3}
  B_1 &= R_{abcd} R_{abcd} R_{efgh} R_{efgh}, \quad&
  B_2 &= R_{abcd} R_{aefg} R_{bcdh} R_{efgh}, \notag
  \\
  B_3 &= R_{abcd} R_{abef} R_{cdgh} R_{efgh}, \quad&
  B_4 &= R_{acbd} R_{aebf} R_{cgdh} R_{egfh}, \notag
  \\
  B_5 &= R_{abcd} R_{aefg} R_{befh} R_{cdgh}, \quad&
  B_6 &= R_{abcd} R_{aefg} R_{bfeh} R_{cdgh}, \label{eq:7base}
  \\
  B_7 &= R_{acbd} R_{aefg} R_{befh} R_{cgdh}. \notag
\end{alignat}
In the main part of this paper we considered the R4 terms
$t_8t_8R^4 - \frac{1}{4!}\epsilon_{11}\epsilon_{11}R^4
= 24 (B_1 - 64B_2 + 2B_3 + 16B_4 - 16 B_5 - 16B_6)$
which is explicitly written in the eq.~(\ref{eq:R4}). 
In order to derive equations of motion, we need to calculate variations of (\ref{eq:7base}).
These are evaluated as
\begin{alignat}{3}
  \dl B_1 &= 4 (\dl R_{abcd}) R_{abcd} R_{efgh} R_{efgh}, \quad
  \dl B_2 = (\dl R_{abcd}) R_{abce} R_{dfgh} R_{efgh}, \notag
  \\
  \dl B_3 &= 4 (\dl R_{abcd}) R_{abef} R_{cdgh} R_{efgh}, \quad
  \dl B_4 = 4 (\dl R_{abcd}) R_{aecg} R_{bfdh} R_{efgh}, \notag
  \\
  \dl B_5 &= 2 (\dl R_{abcd}) R_{abeg} R_{cfeh} R_{dfgh} + 2 (\dl R_{abcd}) R_{efag} R_{efch} R_{gbhd}, 
  \label{eq:7basevar}
  \\
  \dl B_6 &= 2 (\dl R_{abcd}) R_{abeg} R_{cfeh} R_{dfgh} + 2 (\dl R_{abcd}) R_{efag} R_{efch} R_{gbhd}
  - 2 (\dl R_{abcd}) R_{abef} R_{cegh} R_{dfgh}, \notag
  \\
  \dl B_7 &= 4 (\dl R_{abcd}) R_{aefg} R_{cefh} R_{gbhd}. \notag
\end{alignat}
By using these results, we evaluated the eq.~(\ref{eq:R4var}) and derived the equations of motion (\ref{eq:MEOM}).

Next let us consider the R4 terms which necessarily depend on the Ricci tensor or the scalar curvature.
Since the procedure for the classification is straightforward,
we employ a Mathematica code. As as result those are classified into 19 terms.
\begin{alignat}{3}
  B_8 &= R_{abcd} R_{abcd} R_{ef} R_{ef}, \quad&
  B_9 &= R_{abcd} R_{abcd} R^2, \quad&
  B_{10} &= R_{abcd} R_{bcdf} R_{ef} R_{ae}, \notag
  \\
  B_{11} &= R_{abcd} R_{aefg} R_{bcdg} R_{ef}, \quad&
  B_{12} &= R_{abcd}R_{bcde} R_{ae} R, \quad&
  B_{13} &= R_{acbd}R_{cedf} R_{ef} R_{ab} , \notag
  \\
  B_{14} &= R_{abcd} R_{abeg} R_{cdfg} R_{ef}, \quad&
  B_{15} &= R_{acbd} R_{aebg} R_{cfdg} R_{ef}, \quad&
  B_{16} &= R_{abcd} R_{abef} R_{cdef} R, \notag
  \\
  B_{17} &= R_{acbd} R_{aebf} R_{cedf} R, \quad&
  B_{18} &= R_{acbd} R_{ab} R_{cd} R, \quad&
  B_{19} &= R_{abcd} R_{cdef} R_{ae} R_{bf}, \label{eq:19base}
  \\
  B_{20} &= R_{acbd} R_{cedf} R_{ae} R_{bf}, \quad&
  B_{21} &= R_{acbd} R_{ae} R_{be} R_{cd}, \quad&
  B_{22} &= R_{ab} R_{ab} R_{cd} R_{cd}, \notag
  \\
  B_{23} &= R_{ab} R_{ab} R^2, \quad&
  B_{24} &= R_{ab} R_{cd} R_{ac} R_{bd}, \quad&
  B_{25} &= R_{ab} R_{ac} R_{bc} R, \notag
  \\
  B_{26} &= R^4. \notag
\end{alignat}
Then the effective action~(\ref{eq:Maction}) is generalized into the form of
\begin{alignat}{3}
  S_{11} = \frac{1}{2\kappa_{11}^2} \int d^{11}x \; e \Big\{ R
  + \gamma\ell_s^{12} \Big(t_8 t_8 R^4 - \frac{1}{4!} \e_{11} \e_{11} R^4 
  + \sum_{n=8}^{26} b_n B_n \Big) \Big\}. \label{eq:Maction2}
\end{alignat}
The coefficients $b_n (n=8,\cdots, 26)$ cannot be determined from the results of
scattering amplitudes in the type IIA superstring theory, since
we can remove or add these terms by appropriate field redefinitions of the metric.
Therefore it is expected that these terms do not affect physical quantities such as the internal energy
of the black 0-brane. We will confirm this in the appendix \ref{sec:app3}.

Let us derive equations of motion for the effective action (\ref{eq:Maction2}).
The variations of 19 terms in (\ref{eq:19base}) are evaluated as
\begin{alignat}{3}
  \dl B_8 &= (\dl R_{abcd}) \big( 2 R_{abcd} R_{ef} R_{ef} 
  + 2 R_{efgh} R_{efgh} R_{ac}\eta_{bd} \big), \notag
  \\
  \dl B_9 &= (\dl R_{abcd}) \big( 2 R_{abcd} R^2 
  + 2 R_{efgh} R_{efgh} \eta_{ac}\eta_{bd} R \big), \notag
  \\
  \dl B_{10} &= (\dl R_{abcd}) \big( R_{ebcd} R_{af} R_{ef}
  + R_{afgh} R_{efgh} R_{ce} \eta_{bd} \big), \notag
  \\
  \dl B_{11} &= (\dl R_{abcd}) \big( - R_{ebcd} R_{afeg} R_{fg}
  - \tfrac{1}{2} R_{aefg} R_{cefg} R_{bd}
  - \tfrac{1}{2} R_{eghi} R_{fghi} R_{eafc} \eta_{bd} \big), \notag
  \\
  \dl B_{12} &= (\dl R_{abcd}) \big( R_{ebcd} R_{ae} R 
  + \tfrac{1}{2} R_{aefg} R_{cefg} \eta_{bd} R
  + \tfrac{1}{2} R_{eghi} R_{fghi} R_{ef} \eta_{ac} \eta_{bd} \big), \notag
  \\
  \dl B_{13} &= (\dl R_{abcd}) \big( 2 R_{ebfd} R_{ac} R_{ef} 
  + R_{agch} R_{egfh} R_{ef} \eta_{bd} \big), \notag
  \\
  \dl B_{14} &= (\dl R_{abcd}) \big( R_{abeg} R_{cdfg} R_{ef} 
  + 2 R_{abef} R_{efgd} R_{cg}
  + R_{efgh} R_{efai} R_{ghci} \eta_{bd} \big), \notag
  \\
  \dl B_{15} &= (\dl R_{abcd}) \big( R_{aecg} R_{bfdg} R_{ef}
  + 2 R_{aecf} R_{egfd} R_{bg}
  + R_{efgh} R_{eagi} R_{fchi} \eta_{bd} \big), \notag
  \\
  \dl B_{16} &= (\dl R_{abcd}) \big( 3 R_{abef} R_{cdef} R 
  + R_{efgh} R_{efij} R_{ghij} \eta_{ac} \eta_{bd} \big), \notag
  \\
  \dl B_{17} &= (\dl R_{abcd}) \big( 3 R_{aecf} R_{bedf} R
  + R_{efgh} R_{eigj} R_{fihj} \eta_{ac} \eta_{bd} \big), \label{eq:19basevar}
  \\
  \dl B_{18} &= (\dl R_{abcd}) \big( R_{ac} R_{bd} R 
  + 2 R_{aecf} R_{ef} \eta_{bd} R 
  + R_{efgh} R_{eg} R_{fh} \eta_{ac} \eta_{bd} \big), \notag
  \\
  \dl B_{19} &= (\dl R_{abcd}) \big( 2 R_{cdef} R_{ae} R_{bf}
  + 2 R_{aegh} R_{cfgh} R_{ef} \eta_{bd} \big), \notag
  \\
  \dl B_{20} &= (\dl R_{abcd}) \big(2 R_{ebfd} R_{ae} R_{cf} 
  + 2 R_{ageh} R_{cgfh} R_{ef} \eta_{bd} \big), \notag
  \\
  \dl B_{21} &= (\dl R_{abcd}) \big( R_{ae} R_{ce} R_{bd}
  + 2 R_{afeg} R_{ce} R_{fg} \eta_{bd}
  + R_{ebfd} R_{eg} R_{fg} \eta_{ac} \big), \notag
  \\
  \dl B_{22} &= 4 (\dl R_{abcd}) R_{ac} R_{ef} R_{ef} \eta_{bd}, \notag
  \\
  \dl B_{23} &= (\dl R_{abcd}) \big( 2 R_{ac} \eta_{bd} R^2
  + 2 R_{ef} R_{ef} \eta_{ac} \eta_{bd} R \big), \notag
  \\
  \dl B_{24} &= 4 (\dl R_{abcd}) R_{ef} R_{ae} R_{cf} \eta_{bd}, \notag
  \\
  \dl B_{25} &= (\dl R_{abcd}) \big(3 R_{ae} R_{ce} \eta_{bd} R 
  + R_{fg} R_{ef} R_{eg} \eta_{ac} \eta_{bd} \big), \notag
  \\
  \dl B_{26} &= 4 (\dl R_{abcd}) \eta_{ac}\eta_{bd} R^3. \notag
\end{alignat}
And as like the eq.~(\ref{eq:X}), we define $Y$ tensor as
\begin{alignat}{3}
  Y_{abcd} &= \frac{1}{2} \big( Y'_{[ab][cd]} + Y'_{[cd][ab]} \big),
  \\
  Y'_{abcd} &= b_8 \big( 2 R_{abcd} R_{ef} R_{ef} + 2 R_{efgh} R_{efgh} R_{ac}\eta_{bd} \big)
  + b_9 \big( 2 R_{abcd} R^2 + 2 R_{efgh} R_{efgh} \eta_{ac}\eta_{bd} R \big) \notag
  \\
  &\quad\,
  + b_{10} \big( R_{ebcd} R_{af} R_{ef} + R_{afgh} R_{efgh} R_{ce} \eta_{bd} \big) \notag
  \\
  &\quad\,
  + b_{11} \big( - R_{ebcd} R_{afeg} R_{fg} - \tfrac{1}{2} R_{aefg} R_{cefg} R_{bd}
  - \tfrac{1}{2} R_{eghi} R_{fghi} R_{eafc} \eta_{bd} \big) \notag
  \\
  &\quad\,
  + b_{12} \big( R_{ebcd} R_{ae} R + \tfrac{1}{2} R_{aefg} R_{cefg} \eta_{bd} R
  + \tfrac{1}{2} R_{eghi} R_{fghi} R_{ef} \eta_{ac} \eta_{bd} \big) \notag
  \\
  &\quad\,
  + b_{13} \big( 2 R_{ebfd} R_{ac} R_{ef} + R_{agch} R_{egfh} R_{ef} \eta_{bd} \big) \notag
  \\
  &\quad\,
  + b_{14} \big( R_{abeg} R_{cdfg} R_{ef} + 2 R_{abef} R_{efgd} R_{cg}
  + R_{efgh} R_{efai} R_{ghci} \eta_{bd} \big) \notag
  \\
  &\quad\,
  + b_{15} \big( R_{aecg} R_{bfdg} R_{ef} + 2 R_{aecf} R_{egfd} R_{bg}
  + R_{efgh} R_{eagi} R_{fchi} \eta_{bd} \big) \notag
  \\
  &\quad\,
  + b_{16} \big( 3 R_{abef} R_{cdef} R + R_{efgh} R_{efij} R_{ghij} \eta_{ac} \eta_{bd} \big) \notag
  \\
  &\quad\,
  + b_{17} \big( 3 R_{aecf} R_{bedf} R + R_{efgh} R_{eigj} R_{fihj} \eta_{ac} \eta_{bd} \big) \notag
  \\
  &\quad\,
  + b_{18} \big( R_{ac} R_{bd} R + 2 R_{aecf} R_{ef} \eta_{bd} R 
  + R_{efgh} R_{eg} R_{fh} \eta_{ac} \eta_{bd} \big) \notag
  \\
  &\quad\,
  + b_{19} \big( 2 R_{cdef} R_{ae} R_{bf} + 2 R_{aegh} R_{cfgh} R_{ef} \eta_{bd} \big) \notag
  \\
  &\quad\,
  + b_{20} \big(2 R_{ebfd} R_{ae} R_{cf} + 2 R_{ageh} R_{cgfh} R_{ef} \eta_{bd} \big) \notag
  \\
  &\quad\,
  + b_{21} \big( R_{ae} R_{ce} R_{bd} + 2 R_{afeg} R_{ce} R_{fg} \eta_{bd}
  + R_{ebfd} R_{eg} R_{fg} \eta_{ac} \big) \notag
  \\
  &\quad\,
  + 4 b_{22} R_{ac} R_{ef} R_{ef} \eta_{bd} 
  + b_{23} \big( 2 R_{ac} \eta_{bd} R^2 + 2 R_{ef} R_{ef} \eta_{ac} \eta_{bd} R \big) \notag
  \\
  &\quad\,
  + 4 b_{24} R_{ef} R_{ae} R_{cf} \eta_{bd} 
  + b_{25} \big(3 R_{ae} R_{ce} \eta_{bd} R + R_{fg} R_{ef} R_{eg} \eta_{ac} \eta_{bd} \big)
  + 4 b_{26} \eta_{ac}\eta_{bd} R^3. \notag
\end{alignat}
Following the similar calculations in the eq.~(\ref{eq:R4var}),
finally we obtain generic equations of motion
\begin{alignat}{3}
  E_{ij} &\equiv R_{ij} - \frac{1}{2} \eta_{ij} R + \gamma\ell_s^{12} \Big\{
  - \frac{1}{2} \eta_{ij} \Big(t_8 t_8 R^4 - \frac{1}{4!} \e_{11} \e_{11} R^4
  + \sum_{n=8}^{26} b_n B_n \Big) \notag
  \\
  &\quad
  + \frac{3}{2} R_{abci} X^{abc}{}_j - \frac{1}{2} R_{abcj}X^{abc}{}_i - 2 D_{(a} D_{b)} X^a{}_{ij}{}^b \notag
  \\
  &\quad
  + \frac{3}{2} R_{abci} Y^{abc}{}_j - \frac{1}{2} R_{abcj}Y^{abc}{}_i - 2 D_{(a} D_{b)} Y^a{}_{ij}{}^b 
  \Big\} = 0. \label{eq:gMEOM}
\end{alignat}
In order to evaluate these equations, we need to insert the values of the Riemann 
tensor (\ref{eq:Riemann}) into the above.
Since the Ricci tensor and the scalar curvature become zero we obtain $B_n = 0$, and
parts of $b_{11}$, $b_{14}$, $b_{15}$, $b_{16}$ and $b_{17}$ in the $Y$ tensor only contribute 
to the above equations of motion.

Below we repeat the similar calculations in the appendix \ref{sec:app2}.
Each component of $Y_{abcd}$ is evaluated as
\begin{alignat}{3}
  Y_{0101} &= \frac{1}{U_0^6 x^{27} \ell_s^{12}} \Big\{ 
  \frac{11907}{2} b_{11} (1+x^7) - 21609 b_{14} (1+x^7) \notag
  \\
  &\qquad\qquad\qquad\,
  - \frac{3087}{2} b_{15} (1+x^7) - 85176 b_{16} - 10458 b_{17} \Big\}, \notag
  \\[0.1cm]
  Y_{0\bar{a}0\bar{a}} &= \frac{1}{U_0^6 x^{27} \ell_s^{12}} \Big\{ \frac{11907}{8} (-1 + 4 x^7) b_{11} 
  + \frac{63}{4} (5 - 1372 x^7) b_{14} \notag
  \\
  &\qquad\qquad\qquad\,
  - \frac{63}{4} (17 + 98 x^7) b_{15} - 85176 b_{16} - 10458 b_{17} \Big\}, \notag
  \\
  Y_{011\natural} &= \frac{\sqrt{x^7-1}}{U_0^6 x^{\frac{47}{2}} \ell_s^{12}}
  \Big\{ - \frac{11907}{2} b_{11} + 21609 b_{14} + \frac{3087}{2} b_{15} \Big\}, \notag
  \\
  Y_{0\bar{a}\bar{a}\natural} &= \frac{\sqrt{x^7-1}}{U_0^6 x^{\frac{47}{2}} \ell_s^{12}}
  \Big\{ - \frac{11907}{2} b_{11} + 21609 b_{14} + \frac{3087}{2} b_{15} \Big\}, \notag
  \\
  Y_{0\natural 0\natural} &= \frac{1}{U_0^6 x^{27} \ell_s^{12}} \Big\{
  \frac{11907}{2} b_{11} - 21609 b_{14} - \frac{3087}{2} b_{15} - 85176 b_{16} - 10458 b_{17} \Big\}
  \\
  Y_{1\natural1\natural} &= \frac{1}{U_0^6 x^{27} \ell_s^{12}}
  \Big\{ - \frac{11907}{2} b_{11} (2 - x^7) + 21609 (2 - x^7) b_{14} \notag
  \\
  &\qquad\qquad\qquad\,
  + \frac{3087}{2} b_{15} (2 - x^7) + 85176 b_{16} + 10458 b_{17} \Big\}, \notag
  \\[0.1cm]
  Y_{1\bar{a}1\bar{a}} &= \frac{1}{U_0^6 x^{27} \ell_s^{12}}
  \Big\{ - \frac{35721}{8} b_{11} + \frac{86121}{4} b_{14} + \frac{7245}{4} b_{15} 
  + 85176 b_{16} + 10458 b_{17} \Big\}, \notag
  \\
  Y_{\bar{a}\natural \bar{a}\natural} &= \frac{1}{U_0^6 x^{27} \ell_s^{12}}
  \Big\{ \frac{11907}{8} b_{11} (-3 + 4 x^7) + \frac{63}{4} b_{14} (1367 - 1372 x^7) \notag
  \\
  &\qquad\qquad\qquad\,
  + \frac{63}{4} b_{15} (115 - 98 x^7) + 85176 b_{16} + 10458 b_{17}, \notag
  \\[0.1cm]
  Y_{\bar{a}\bar{b}\bar{a}\bar{b}} &= \frac{1}{U_0^6 x^{27} \ell_s^{12}}
  \Big\{ \frac{11907}{4} b_{11} - \frac{315}{2} b_{14} + \frac{1071}{2} b_{15} 
  + 85176 b_{16} + 10458 b_{17} \Big\}, \notag
\end{alignat}
where $\bar{a},\bar{b}=2,\cdots,9$.
By using these results it is possible to evaluate the higher derivative terms 
which depend on the Y tensor in the eq.~(\ref{eq:gMEOM}).
The $RY$ terms are calculated as
\begin{alignat}{3}
  R_{abc0} Y^{abc}{}_0 &= \frac{1}{U_0^8 x^{29} \ell_s^{16}} 
  \Big\{ 416745 b_{11} - 1214514 b_{14} - 71442 b_{15} \Big\}, \notag
  \\
  R_{abc1} Y^{abc}{}_1 &= \frac{1}{U_0^8 x^{36} \ell_s^{16}} 
  \Big\{ -416745 b_{11} + 1214514 b_{14} + 71442 b_{15} \Big\}, \notag
  \\
  R_{abc\natural} Y^{abc}{}_\natural &= \frac{x^7-1}{U_0^8 x^{36} \ell_s^{16}}
  \Big\{ 416745 b_{11} - 1214514 b_{14} - 71442 b_{15} \Big\},
  \\
  R_{abc\bar{a}} Y^{abc}{}_{\bar{b}} &= \frac{1}{4 U_0^8 x^{36} \ell_s^{16}} \delta_{\bar{a}\bar{b}}
  \Big\{ 416745 b_{11} - 1214514 b_{14} - 71442 b_{15} \Big\}, \notag
  \\
  R_{abc0} Y^{abc}{}_\natural &= R_{abc\natural} Y^{abc}{}_{0} = 
  \frac{\sqrt{x^7-1}}{U_0^8 x^{\frac{65}{2}} \ell_s^{16}}
  \Big\{ 416745 b_{11} - 1214514 b_{14} - 71442 b_{15} \Big\}, \notag
\end{alignat}
and $DDY$ terms become
\begin{alignat}{3}
  D_{(a} D_{b)} Y^a{}_{00}{}^b &= \frac{1701}{U_0^8 x^{36} \ell_s^{16}}
  \Big\{ - \frac{7}{2} (-459 - 235 x^7 + 540 x^{14}) b_{11} \notag
  \\
  &\quad\,
  + (-6507 - 2397 x^7 + 6860 x^{14}) b_{14} + \frac{1}{2} (-999 - 282 x^7 + 980 x^{14}) b_{15} \notag
  \\
  &\quad\,
  + (-36504 + 31772 x^7) b_{16} + (-4482 + 3901 x^7) b_{17} \Big\}, \notag
  \\
  D_{(a} D_{b)} Y^a{}_{11}{}^b &= \frac{1701}{U_0^8 x^{36} \ell_s^{16}}
  \Big\{ -7 (31 + 46 x^7) b_{11} + 4 (6 + 505 x^7) b_{14} \notag
  \\
  &\quad\,
  + \frac{1}{2} (-75 + 376 x^7) b_{15} + 676 (-9 + 16 x^7) b_{16} + 83 (-9 + 16 x^7) b_{17} \Big\}, \notag
  \\
  D_{(a} D_{b)} Y^a{}_{\natural\natural}{}^b &= 
  \frac{1701}{U_0^8 x^{36} \ell_s^{16}}
  \Big\{ - \frac{7}{2} (1034 - 1455 x^7 + 540 x^{14}) b_{11} \notag
  \\
  &\quad\,
  + (13724 - 18897 x^7 + 6860 x^{14}) b_{14} + \frac{1}{2} (2021 - 2742 x^7 + 980 x^{14}) b_{15} \notag
  \\
  &\quad\,
  + 676 (47 - 33 x^7) b_{16} + 83 (47 - 33 x^7) b_{17} \Big\}, 
  \\
  D_{(a} D_{b)} Y^a{}_{\bar{a}\bar{b}}{}^b &= \frac{-4 + 3 x^7}{U_0^8 x^{36} \ell_s^{16}} 
  \delta_{\bar{a}\bar{b}}
  \Big\{ \frac{1917027}{4} b_{11} - \frac{6013035}{2} b_{14} - \frac{559629}{2} b_{15} \notag
  \\
  &\qquad\qquad\qquad\quad\,
  - 16098264 b_{16} - 1976562 b_{17} \Big\}, \notag
  \\
  D_{(a} D_{b)} Y^a{}_{0\natural}{}^b &= D_{(a} D_{b)} Y^a{}_{\natural 0}{}^b \notag
  \\
  &= \frac{\sqrt{x^7-1}}{U_0^8 x^{\frac{65}{2}} \ell_s^{16}}
  \Big\{ \frac{59535}{2} (115 - 108 x^7) b_{11} - 11907 (1031 - 980 x^7) b_{14} \notag
  \\
  &\quad\,
  - 11907 (73 - 70 x^7) b_{15} + 8049132 b_{16} + 988281 b_{17} \Big\}. \notag
\end{alignat}
As mentioned before, only $b_{11}$, $b_{14}$, $b_{15}$, $b_{16}$ and $b_{17}$ appeared in the calculations.

By using the ansatz (\ref{eq:Mansatz}) and
inserting values of $X$ and $Y$ tensors into the equations of motion (\ref{eq:gMEOM}), 
we obtain five independent equations with parameters $b_{11}$, $b_{14}$, $b_{15}$, $b_{16}$ and $b_{17}$.
\begin{alignat}{3}
  E_1 &= - 63 x^{34} f_1 - 9 x^{35} f_1' 
  - 49 x^{41} h_1 + 49 x^{34}(1- x^7) h_2 + 23 x^{35}(1- x^7) h_2' + 2 x^{36}(1- x^7) h_2'' \notag
  \\
  &\quad\,
  + 98 x^{41} h_3 + 7 x^{42} h_3' - 63402393600 x^{14} + 70230343680 x^7 + 1062512640 \notag
  \\
  &\quad\,
  + (25719120 b_{11} - 93350880 b_{14} - 6667920 b_{15}) x^{14} \label{eq:1'}
  \\
  &\quad\,
  + (-9525600 b_{11} + 27760320 b_{14} + 1632960 b_{15} - 432353376 b_{16} - 53084808 b_{17}) x^7 \notag
  \\
  &\quad\,
  - 21861252 b_{11} + 88547256 b_{14} + 6797196 b_{15} + 496746432 b_{16} + 60991056 b_{17}
  = 0, \notag
  \\[0.2cm]
  E_2 &= 63 x^{34} f_1 + 9 x^{35} f_1' + 7 x^{34}(9 - 2 x^{7}) h_1 + 9 x^{35} (1 - x^{7}) h_1' 
  - 112 x^{34} (1 - x^{7}) h_2 \notag
  \\
  &\quad\,
  - 16 x^{35}(1 - x^{7}) h_2' - 98 x^{41} h_3 - 7 x^{42} h_3' - 2159861760 x^7- 5730600960 \label{eq:2'}
  \\
  &\quad\,
  + (4381776 b_{11} - 27488160 b_{14} - 2558304 b_{15} - 147184128 b_{16} - 18071424 b_{17}) x^7 \notag
  \\
  &\quad\,
  + 1285956 b_{11} + 4531464 b_{14} + 796068 b_{15} + 82791072 b_{16} + 10165176 b_{17}
  = 0, \notag
  \\[0.2cm]
  E_3 &= 133 x^{34} f_1 + 35 x^{35} f_1' + 2 x^{36} f_1'' 
  + 28 x^{34} (3 - 10 x^{7}) h_1 + 7 x^{35} (4 - 7 x^{7}) h_1' + 2 x^{36} (1 - x^{7}) h_1'' \notag
  \\
  &\quad\,
  - 7 x^{34} (5 - 26 x^{7}) h_2 - 21 x^{35} (1 - 2 x^{7}) h_2' - 2 x^{36} (1 - x^{7}) h_2'' 
  + 98 x^{41} h_3 + 7 x^{42} h_3' \notag
  \\
  &\quad\,
  + 5669637120 x^7 - 8626383360 \label{eq:3'}
  \\
  &\quad\,
  + (-11502162 b_{11} + 72156420 b_{14} + 6715548 b_{15} + 386358336 b_{16} + 47437488 b_{17}) x^7 \notag
  \\
  &\quad\,
  + 15752961 b_{11} - 97423074 b_{14} - 9025506 b_{15} - 515144448 b_{16} - 63249984 b_{17} 
  = 0, \notag
  \\[0.2cm]
  E_4 &= 259 x^{34} f_1 + 53 x^{35} f_1' + 2 x^{36} f_1'' 
  + 147 x^{34}(1 - 3 x^{7}) h_1 + x^{35}(37 -58 x^{7}) h_1' \notag
  \\
  &\quad\,
  + 2 x^{36}(1 - x^{7}) h_1'' 
  + 147 x^{41} h_2 + 21 x^{42} h_2'
  + 294 x^{41} h_3 + 21 x^{42} h_3' \notag
  \\
  &\quad\,
  - 63402393600 x^{14} + 133632737280 x^7 - 71292856320 \label{eq:4'}
  \\
  &\quad\,
  + x^{14} (25719120 b_{11} - 93350880 b_{14} - 6667920 b_{15}) \notag
  \\
  &\quad\,
  + x^7 (-67631760 b_{11} + 252292320 b_{14} + 18370800 b_{15} + 303567264 b_{16} + 37272312 b_{17}) \notag
  \\
  &\quad
  + 47580372 b_{11} - 181898136 b_{14} - 13465116 b_{15} - 432353376 b_{16} - 53084808 b_{17} 
  = 0, \notag
  \\[0.2cm]
  E_5 &= 49 x^{34} h_1 + 7 x^{35} h_1' + 49 x^{34} h_2 - x^{35} h_2' - x^{36} h_2'' 
  - 98 x^{34} h_3 - 22 x^{35} h_3' - x^{36} h_3'' \notag
  \\
  &\quad\,
  - 63402393600 x^7 + 70230343680 \label{eq:5'}
  \\
  &\quad\,
  + x^7 (25719120 b_{11} - 93350880 b_{14} - 6667920 b_{15}) \notag
  \\
  &\quad\,
  - 25719120 b_{11} + 93350880 b_{14} + 6667920 b_{15} - 64393056 b_{16} - 7906248 b_{17}
  = 0. \notag
\end{alignat}
Here we defined $E_1 = 4U_0^8 \ell_s^4 x^{36}\gamma^{-1} E_{00}$,
$E_2 = 4U_0^8 \ell_s^4 x^{36}\gamma^{-1} E_{11}$, $E_3 = 4U_0^8 \ell_s^4 x^{36}\gamma^{-1} E_{22}$,
$E_4 = 4U_0^8 \ell_s^4 x^{36}\gamma^{-1} E_{\natural\natural}$ and
$E_5 = 4U_0^8 \ell_s^4 x^{\frac{65}{2}}(-1+x^7)^{-\frac{1}{2}}\gamma^{-1} E_{0\natural}$.

The equations (\ref{eq:1'})--(\ref{eq:5'}) can be solved by following the details in the section \ref{sec:sol}.
And the final form of the solution becomes
\begin{alignat}{3}
  h_1 &= \Big(-\frac{440559}{4} b_{11} + \frac{768775}{2} b_{14}
  + \frac{53333}{2} b_{15} + 927472 \,b_{16} 
  + 113876 b_{17} + \frac{1302501760}{9} \Big) \frac{1}{x^{34}} \notag
  \\
  &\quad\,
  + \big( 23814 \,b_{11} - 86436 \,b_{14} - 6174 \,b_{15} - 57462496 \big) \frac{1}{x^{27}} 
  +\frac{12051648}{13 x^{20}}-\frac{4782400}{13x^{13}} \notag
  \\
  &\quad\,
  - \frac{3747840}{x^7} + \frac{4099200}{x^6} - \frac{1639680 (x-1)}{(x^7-1)} 
  + 117120 \Big( 18 - \frac{23}{x^7} \Big) I(x), \notag
  \\[0.2cm]
  h_2 &= \Big(-\frac{11907}{4} b_{11} + \frac{315}{2} b_{14} - \frac{1071}{2} b_{15}
  - 170352 \,b_{16} - 20916 \,b_{17} + 19160960 \Big) \frac{1}{x^{34}} \notag
  \\
  &\quad\,
  + \big( 23814 \,b_{11} - 86436 \,b_{14} - 6174 \,b_{15} - 58528288 \big) \frac{1}{x^{27}}
  +\frac{2213568}{13 x^{20}} -\frac{1229760}{13x^{13}} \notag
  \\
  &\quad\,
  - \frac{2108160}{x^7} + \frac{2459520}{x^6} 
  + 1054080 \Big( 2 - \frac{1}{x^7} \Big) I(x), \label{eq:gsol}
  \\[0.2cm]
  h_3 &= \Big(-\frac{11907}{4} b_{11} + \frac{76027}{2} b_{14} 
  + \frac{8225}{2} b_{15} - 94640 \,b_{16} - 11620 \,b_{17} 
  + \frac{361110400}{9} \Big) \frac{1}{x^{34}} \notag
  \\
  &\quad\,
  + \big( 23814 \,b_{11} - 86436 \,b_{14} - 6174 \,b_{15} - 59840032 \big) \frac{1}{x^{27}}
  - \frac{24021312}{13x^{20}} \notag 
  \\
  &\quad\, 
  - \frac{58072000}{13x^{13}} - \frac{2108160}{x^7} + \frac{2459520}{x^6} 
  + 117120 \Big(18 - \frac{41}{x^7} \Big) I(x), \notag
  \\[0.2cm]
  f_1 &= \Big( \frac{440559}{4} b_{11} - \frac{730919}{2} b_{14} 
  - \frac{48685}{2} b_{15} - 889616 \,b_{16} 
  - 109228 \,b_{17} - \frac{1208170880}{9} \Big) \frac{1}{x^{34}} \notag
  \\
  &\quad\,
  + \big( -130977 \,b_{11} + 432810 \,b_{14} 
  + 28728 \,b_{15} + 1022112 \,b_{16} + 125496 \,b_{17} + 161405664 \big) \frac{1}{x^{27}} \notag
  \\
  &\quad\,
  + \frac{5738880}{13x^{20}} + \frac{956480}{x^{13}} \!+\! \frac{819840}{x^7} I(x). \notag
\end{alignat}
The function $I(x)$ is given by the eq.~(\ref{eq:I}) and integral constants are determined so as to satisfy that
$h_i(1)$ are finite and $h_i(x), f_1(x) \sim \mathcal{O}(x^{-8})$ when $x$ goes to the infinity.
Notice that $b_{11}$, $b_{14}$, $b_{15}$, $b_{16}$ and $b_{17}$ only appear in the coefficients of $x^{-27}$
and $x^{-34}$.
The solution is reliable up to $\mathcal{O}(\epsilon^2)$.

\section{Thermodynamics of Black 0-Brane with Generic $R^4$ Terms}\label{sec:app3}

In this appendix, we examine thermodynamics of the quantum near horizon geometry of the black 0-brane (\ref{eq:gsol})
by following the arguments in the section \ref{sec:thermo}.
Although the solution is modified, the results obtained until the eq.~(\ref{eq:Wald}) do not change.
Since the effective action is modified as in the eq.~(\ref{eq:Maction2}), 
the eq.~(\ref{eq:varS}) should be replaced with
\begin{alignat}{3}
  \frac{\partial S_{11}}{\partial R_{\mu\nu\rho\sigma}} &=
  \frac{1}{2\kappa_{11}^2} \big\{ g^{\mu[\rho}g^{\sigma]\nu} 
  + \gamma \ell_s^{12} (X^{\mu\nu\rho\sigma}+Y^{\mu\nu\rho\sigma}) \big\}. \label{eq:varS2}
\end{alignat}
The entropy of the quantum near horizon geometry of the black 0-brane is evaluated as
\begin{alignat}{3}
  S &= \frac{4\pi}{2\kappa_{11}^2} \int_\text{H} d\Omega_8 dz \sqrt{h}
  \Big( 1 - \frac{1}{2} \gamma \ell_s^{12} (X^{\mu\nu\rho\sigma} + Y^{\mu\nu\rho\sigma})
  N_{\mu\nu} N_{\rho\sigma} \Big) \notag
  \\
  &= \frac{4\pi}{2\kappa_{11}^2} \int_\text{H} d\Omega_8 dz \sqrt{h}
  \big( 1 - 2 \gamma \ell_s^{20} U_0^2 H_1^{-1} (X^{txtx}+Y^{txtx}) \big) \notag
  \\
  &= \frac{4\pi}{2\kappa_{11}^2} \int_\text{H} d\Omega_8 dz \sqrt{h}
  \Big\{ 1 + \epsilon \,{U}_0^{-6} \big( 40642560 \notag
  \\
  &\qquad\qquad\qquad
  - 23814 b_{11} + 86436 b_{14} + 6174 b_{15} 
  + 170352 b_{16} + 20916 b_{17} \big) \Big\} \notag
  \\
  &= \frac{4}{49} a_1 N^2 \tilde{U}_0^\frac{9}{2}
  \Big\{ 1 + \epsilon \Big( - \frac{9}{14} f_1(1) + \frac{1}{2} h_2(1) + 40642560 \notag
  \\
  &\qquad\qquad\qquad
  - 23814 b_{11} + 86436 b_{14} + 6174 b_{15} 
  + 170352 b_{16} + 20916 b_{17} \Big) \tilde{U}_0^{-6} \Big\} \notag
  \\
  &= \frac{4}{49} a_1^{-\frac{4}{5}} N^2 \tilde{T}^\frac{9}{5}
  \Big\{ 1 + \epsilon a_1^{\frac{12}{5}}\Big( - \frac{9}{5} f_1(1) - \frac{9}{35} f'_1(1)
  + \frac{9}{10} h_1(1) + \frac{1}{2} h_2(1) + 40642560 \notag
  \\
  &\qquad\qquad\qquad
  - 23814 b_{11} + 86436 b_{14} + 6174 b_{15} 
  + 170352 b_{16} + 20916 b_{17} \Big) 
  \tilde{T}^{-\frac{12}{5}} \Big\} \notag
  \\[0.1cm]
  &= a_3 N^2 \tilde{T}^\frac{9}{5}
  \big( 1 + \epsilon \, a_5 \tilde{T}^{-\frac{12}{5}} \big). \label{eq:entropy2}
\end{alignat}
Notice that $f_1(1)$, $f'_1(1)$, $h_1(1)$ and $h_2(1)$ depend on 
$b_{11}$, $b_{14}$, $b_{15}$, $b_{16}$ and $b_{17}$.
The value of $a_3$ is given in the section \ref{sec:thermo}, and $a_5$ is given by
\begin{alignat}{3}
  a_5 &= a_1^{\frac{12}{5}} \Big( - \frac{9}{5}f_1(1) - \frac{9}{35}f'_1(1) 
  + \frac{9}{10}h_1(1) + \frac{1}{2}h_1(1) + 40642560 \notag
  \\
  &\qquad\quad\;\,
  - 23814 b_{11} + 86436 b_{14} + 6174 b_{15} 
  + 170352 b_{16} + 20916 b_{17} \Big).
\end{alignat}
It seems that $a_5$ depends on $b_{11}$, $b_{14}$, $b_{15}$, $b_{16}$ and $b_{17}$.
The explicit calculation, however, shows that $a_5 = a_4$ and
the result does not depend on the ambiguities of the effective action.
Thus the physical quantities of the black 0-brane are free from the ambiguities and uniquely determined.

\end{document}